\documentclass[twocolumn,times]{aastex701}
\usepackage{graphicx} 

\usepackage{xcolor}

\newcommand{\lya}{$\mathrm{Ly}\alpha$}
\newcommand{\cii}{[C\textsc{\,ii}]}
\newcommand{\hi}{H\textsc{\,i}}
\newcommand{\limstacker}{\texttt{lim\_stacker}}
\newcommand{\lprimeco}{$L'_{\rm CO}$}
\usepackage{tikz}
\usepackage{newtx}
\usepackage{lineno}
\usepackage{hyperref}
\usepackage{cleveref}
\usepackage{booktabs}

\begin{document}
\nolinenumbers
\title{Clustering Confuses Spectro-photometry: An Investigation of 2D and 3D Forced Profile Matching for Stacking Line-intensity Mapping Data on Source Catalogues}
\author{Ella M.~Mansfield}
\affiliation{Department of Physics, Cornell University, Ithaca NY 14853, USA}
\email{emm362@cornell.edu}
\author[0000-0002-5223-8315]{Delaney A.~Dunne}
\affiliation{California Institute of Technology, 1200 E. California Blvd., Pasadena, CA 91125, USA}
\email{ddunne@astro.caltech.edu}
\author[0000-0003-2618-6504]{Dongwoo T.~Chung}
\affiliation{Department of Astronomy, Cornell University, Ithaca NY 14853, USA}
\email{dongwooc@cornell.edu}
\date{December 2025}


\begin{abstract}
    Line-intensity mapping (LIM) is an emerging observational technique that is used to observe the universe on large scales at low resolution through spectral line emission. Stacking analyses coadd cutouts of LIM data on positions of known signal emitters, robustly detecting signal otherwise hidden in a noisy map. In this article, we present two augmentations of a stacking pipeline, both aiming to refine the sensitivity of the stack by assuming a specific observed signal shape in 2D spatial axes or 3D spatial and spectral axes, as well as stacking on source coordinates more precise than the resolution of the LIM data cube. We test these methods on a series of simplistic and complex simulations mimicking observations with the CO Mapping Array Project (COMAP) Pathfinder. We find that these fitting methods provide up to a 25\% advantage in detection significance over the original stack method in realistic COMAP-like simulations. We also find that the optimal fitting profile, given our CO line model and galaxy model, is larger than the $5'$ width of the COMAP beam and takes on a Lorentzian shape in the spectral dimension. Our findings suggest a nuanced dependence of the optimal profile size and shape on the LIM signal itself, including redshift-space clustering and fingers-of-God effects that depend on the tracer luminosity function and bias.
\end{abstract}
\section{Introduction}
\label{sec:intro}
\begin{figure*}[t!!]
    \centering
    \begin{tikzpicture}
    \node () at (0,0) {\includegraphics[width=0.96\linewidth]{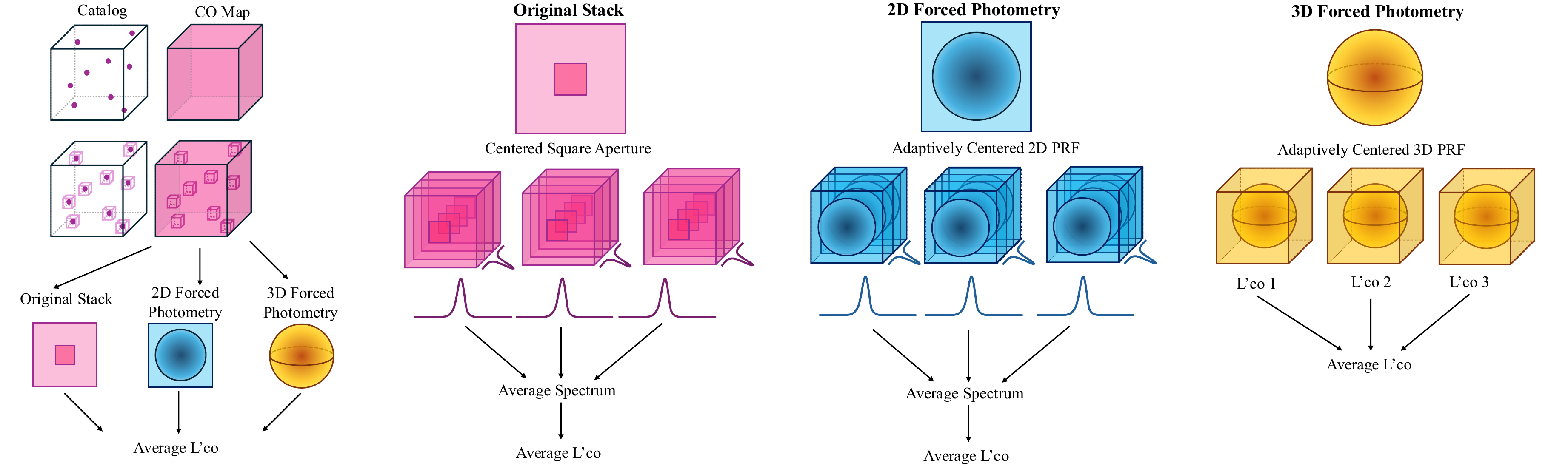}};
    \draw[dotted] (-4.586,-2.86) -- (-4.586,2.86);
    \node[fill=white,font=\tiny] () at (-6.642,-2.486) {Average \lprimeco{}};
    \node[fill=white,font=\tiny] () at (-2.486,-2.542) {Average \lprimeco{}};
    \node[fill=white,font=\tiny] () at (2.,-2.542) {Average \lprimeco{}};
    \node[fill=white,font=\tiny] () at (6.486,-1.542) {Average \lprimeco{}};
    \node[fill=white,font=\tiny] () at (5.286,-0.586) {\lprimeco{} (1)};    
    \node[fill=white,font=\tiny] () at (6.486,-0.586) {\lprimeco{} (2)};
    \node[fill=white,font=\tiny] () at (7.686,-0.586) {\lprimeco{} (3)};
    \end{tikzpicture}
    \caption{A depiction of the stacking analysis and its three different stacking methods.}
    \label{fig:StackSynopsis Placeholder}
\end{figure*}
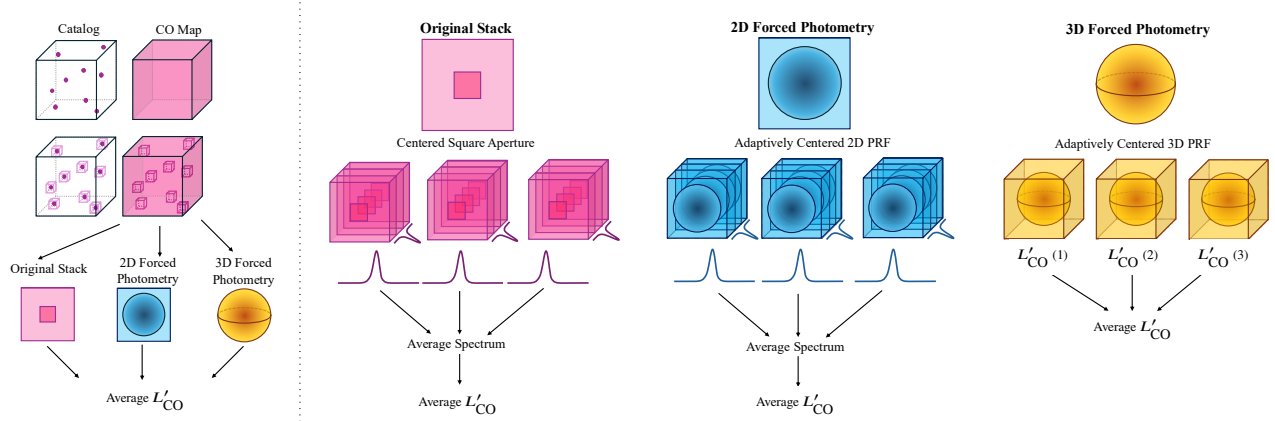
Line intensity mapping (LIM) is an observational technique that considers the aggregate spectral line emission of all sources in an observation field across large patches of the sky~\citep[see reviews by, e.g.:][]{Kovetz17,BernalKovetz22}. LIM is primarily used to obtain low-resolution maps of the universe at different redshifts to understand its structure on very large scales. 

The Carbon monOxide Mapping Array Project (COMAP) Pathfinder \citep{Cleary_2022} is an example of an experiment using this technique to trace molecular carbon monoxide (CO) transitions. The rotational CO(1--0) transition at rest-frame emission frequency 115.27 GHz is a valuable tracer of star formation. It can trace cool molecular clouds in star-forming regions most directly~\citep[for a review, see][]{CW13} compared to other typical LIM target lines (e.g. H$\alpha$, \cii{}, $21$\,cm \hi{}), which only trace them indirectly or not at all. The COMAP Pathfinder focuses on mapping this transition between redshifts $2.4$ to $3.4$, corresponding to a frequency range of 26-34 GHz. For a more complete description of the instrumentation of the COMAP Pathfinder, see \cite{Lamb_2022}. 

COMAP is an advanced, dedicated facility for star-formation LIM, a subfield where similar experiments are still in an era of commissioning---e.g., CONCERTO~\citep{Desert25} and TIME~\citep{Yang25} for \cii{}---and targeting first detections. Early detections, even of the brightest peaks of the line-intensity signal, present a significant challenge for these experiments. The low resolution instrumentation used in LIM leads to signal from unresolved sources being attenuated across a single pixel, producing a lower signal to noise ratio (SNR) and making detections more difficult. These reasons strongly motivate the optimization of analysis methods to accelerate the initial detection of a signal or provide more accurate upper limits on the luminosities of undetected signals.

In this article, we thoroughly analyze two augmentations of a COMAP data analysis pipeline first presented in \cite{dunne2024comapearlyscienceviii} used to `stack' the COMAP data cube on sources in the eBOSS quasar sample \citep{dawson2016_ebossoverview}. Combining LIM data with catalogues of known emitters is a robust way to make more confident detections, as shown repeatedly in the context of low-redshift 21\,cm \hi{} LIM~\citep{Chang10,Masui13,CHIME2023,Chen25MeerKLASS} and most recently in $z\sim3$ Lyman-$\alpha$ (\lya) LIM using the Hobby--Eberly Telescope Dark Energy eXperiment~\citep[HETDEX;][]{Niemeyer25}. The augmentations to the stacking pipeline presented in this work assume specific signal profiles in 2D and 3D with the goal of lowering the uncertainty around recovered line luminosity values, thus increasing the SNR with which we recover said line luminosity. 

In addition to incorporating prior information about the stack signal, the assumption of such profiles also allows for sub-voxel positioning of those profiles in fitting. This may provide an advantage over simply coadding cutouts at the voxel level, as the coordinates of the CO sources stacked on are typically known to an accuracy far greater than the resolution of COMAP. For instance, source coordinates in the HETDEX catalogue of $\rm{Ly}\alpha$ emitters---a primary cross-correlation target for COMAP~\citep{comap_xcorr}---are known to within a few arcseconds~\citep{Mentuch_Cooper_2023}. This is about 40$\times$ smaller than COMAP pixels, which are $2'$ wide to Nyquist-sample the beam FWHM of 4.5$'$ at 30\,GHz~\citep{Lamb_2022}. In the spectral axis, the COMAP data are binned to a 31.25\,MHz resolution for scientific analysis, for a resolving power of $\nu/\delta\nu\sim1000$ much better matched to the HETDEX redshift accuracy. 

By supplying the stacking analysis of LIM data with information such as signal profile models in the form of point response functions (PRF; thus weighting most heavily the voxels which contain the most signal), and sub-pixel/sub-voxel coordinates (thus concentrating the signal with respect to the noise), we aim to achieve more accurate and confident signal recovery in contexts like the COMAP Pathfinder. The work is extremely timely as the latest results from COMAP Pathfinder stacks~\citep{dunne2025_comaps2iv} on external catalogs are beginning to disfavor an increasing volume of the parameter space for the luminosity function of CO emission.

We devise and test our augmented profile-based stacking pipelines in the limit of uniformly distributed sources with completely non-overlapping profiles, and on realistically clustered simulations of CO emitters. In doing so, this work aims to answer the following questions:
\begin{itemize}
    \item Can the forced fitting of 2D and 3D profiles at the sub-pixel level enhance the accuracy and significance of existing LIM stacking methods?
    \item Do the optimal parameters for these profiles depend solely on the instrument and survey design, or do they also depend on the LIM signal itself?
\end{itemize}

We structure the paper as follows: \hyperref[sec:stackingmethods]{Section 2} provides a description of each of the three stacking methods presented in this article and how they derive a final luminosity and associated uncertainty. \hyperref[sec:sanitycheck]{Section 3} and \hyperref[sec:realisticsims]{Section 4} discuss simulated tests discerning the behavior of the two forced stacking methods against the traditional stacking method. \hyperref[sec:discussion]{Section 5} discusses the findings before we consider future directions in the concluding~\autoref{sec:conclusions}.

Where relevant, we assume a flat $\Lambda$CDM cosmology with $\Omega_m=0.286$, $\Omega_b=0.047$, $H_0=100h$ km s$^{-1}$ Mpc$^{-1}$ with $h=0.7$, and $\sigma_8=0.82$, for consistency with previous COMAP studies.

\section{Stacking Methods}\label{sec:stackingmethods}
\subsection{The Original Stack}
\cite{dunne2024comapearlyscienceviii} describes a stacking pipeline (hereafter referred to as \limstacker{}) used to detect a signal from data obtained from the COMAP Season 1 data release \citep{Cleary_2022}. Throughout this work, we will refer to the output of this pipeline as the `original' stack, and use it as a baseline for comparing the stacking methods developed in this article. We thus describe its methodology here as a point of comparison.

Each of the stacking pipelines we describe here use two sources of data -- the 3D LIM data cube, and a reference catalog of galaxies with known three dimensional positions (right ascension and declination in the spatial axes, and redshift in the spectral axis). In \cite{dunne2024comapearlyscienceviii}, the reference catalog is the eBOSS quasar sample \citep{Dawson_2016}. \limstacker{} uses the reference catalog to create three-dimensional cutouts (cubelets) from the COMAP data cube around each catalog source, with cubelet dimensions of $42'\times42'\times1.28$ GHz, centered on the voxel containing the source. These are purposely much larger than the spatial/spectral extent of the expected signal, in order to inspect the final stacked cube for large-scale structure.

 Cubelets are stacked by coadding subsequent cubelets into a running average cubelet, using inverse variance weighting. Voxels (3D pixels) in each new cubelet are averaged into the corresponding voxels in the running average cubelet, and the end result is a single average cubelet created from each cutout. A square aperture including the central $N_\mathrm{spax}\times N_\mathrm{spax}$ spatial pixels (where $N_\mathrm{spax}$ is the side length of the aperture in spatial pixels) of this average cubelet is summed over in each spectral bin, resulting in an average `spectrum' of luminosity along the spectral axis. The 2D aperture over which these voxels are summed is determined based on the assumption that the spatial distribution of the signal will be primarily concentrated towards the center of the cubelet. The actual sources are unresolved at the resolution of the pathfinder instrument, so it is assumed that at the least the telescope beam spreads the signal outside of the central voxel. \cite{dunne2025_stacktheory} found that, due to astrophysical clustering, the signal is spread beyond the size of the beam, extending several arcminutes beyond the beam FWHM. For their signal models (the same used in this work), \cite{dunne2025_stacktheory} found that aperture side lengths between $N_\mathrm{spax} = 5$ and $N_\mathrm{spax} = 9$ maximized the stack SNR. No other significant assumptions are made about the shape of the source. The final luminosity value and associated uncertainty are obtained by summing the luminosities in the central $N_\mathrm{chan}$ channels (frequency bins) of the average spectrum and then calculating the error associated with that sum, either by propagating the RMS uncertainty in each voxel of the COMAP data or by using a bootstrapping technique (described in Section \ref{sec:bootstraps}).

When we refer to this original stacking technique, we default to an aperture size of $N_\mathrm{spax}\times N_\mathrm{spax} \times N_\mathrm{chan} = 5 \times 5 \times 3$ throughout this work, meaning we integrate flux over the central $5\times5$ pixels to obtain a spectrum and then sum over the central 3 channels to obtain a stack luminosity.

\subsection{Forced Photometry with a 2D PRF}
In this work, we introduce the application of 2D forced photometry to LIM stacking, refining the idea of choosing a wider aperture to account for signal spread. This method assumes the signal shape in the right ascension and declination axes to be Gaussian. Because the sources targeted by COMAP are unresolved at the resolution of the instrument, the na\"{i}ve expectation would be that any signal shape in these axes is  governed by the shape of the telescope beam itself and effectively spread outside the pixel from which it originated. In the case of COMAP, the beam profile may be approximated by a single, circular, Gaussian function, as the beam shows excellent sidelobes (20 dB below main beam for the central feed) and only a 4\% variation in FWHM with frequency up or down relative to the median value across the band~\citep{Lamb_2022}. However, at sufficiently low resolution, the confusion of different sources may motivate anticipation of an extended component in the stack from pseudo-resolved clustering, as described by~\cite{dunne2025_stacktheory}.

We use the \texttt{photutils} implementation of forced photometry to extract a signal from each frequency channel of the coadded cubelet based on a 2D Gaussian PRF, instead of simply summing across the central pixels in a square aperture. A PRF model combines the pixel window of the LIM data cube grid with the point spread function (PSF) model in the limit of infinite oversampling. The width of the PRF can be chosen to match that of the actual COMAP beam convolved with its pixel window function, but can equally be chosen to match an expected extended component. For this work, we keep the PRF width consistent for all cutouts in a given stack, and the PRF is calculated based on the exact coordinates of each source to fit COMAP data at a sub-pixel level. The $L'_{\rm CO}$ errors produced by this extraction method are estimated based on a least-squares fit.

We calculate a luminosity spectrum and its associated errors for each cutout by performing the forced photometry in each frequency channel in a cubelet individually. The cubelets are then stacked per-voxel in the same way as the original stack based on an inverse variance weighting scheme, with the main difference being that each cubelet now has a different associated spectrum achieved with 2D photometry than they would in the original stack. The values along the \lprimeco{} spectrum of the final stacked cube are summed together for a final average \lprimeco{} value across all cubelets. Here, we only sum values within the central $N_\mathrm{chan}=3$ channels, mirroring the parameters chosen for the original stack.

\subsection{Forced Spectro-photometry with a 3D PRF}
3D forced spectro-photometry stretches the ideas from 2D forced photometry into all three dimensions. The function assumes the signal shape in all three axes and fits a three-dimensional spatial-spectral PRF to the data based on this this assumption. This PRF can be of any form, but here we consider either a 3D Gaussian or a 2D Gaussian in space combined with a 1D Lorentzian spectral profile. The 3D Gaussian best represents the emission profile of individual sources in our mock CO maps, which use the simulation methodology outlined in~\cite{dunne2025_stacktheory}, while the Lorentzian spectral profile best represents the average stack spectrum found in simulations by~\cite{dunne2025_stacktheory}. This 3D PRF pipeline uses a Levenberg--Marquandt method\footnote{We specifically use the \texttt{scipy} \citep{2020SciPy-NMeth} library's implementation through \texttt{curve\_fit}.} to then fit the amplitude of this function to the data. With the PRF model normalized such that the fitted amplitude corresponds to \lprimeco{}, we record the \lprimeco{} of that cubelet along with the covariance output by the fitting method.

Note that as we progress from the original stack to 2D and then 3D PRF stacks, we retain less and less information per cutout, using the PRF to reduce 2D or 3D information into a single number. 2D forced photometry functions more similarly to the original stack in the sense of extracting a spectrum from each cubelet, with the final value taken via a sum over an average spectrum of all cubelets. 3D forced photometry bypasses the calculation of a \lprimeco{} spectrum entirely, as fitting a 3D PRF to each cubelet compresses all of the information in that cubelet into a single total \lprimeco{} value. These luminosity values are averaged together at the end of the pipeline based on an inverse variance weighting scheme.

\begin{deluxetable*}{p{3.586cm}lp{4.286cm}p{4.286cm}}
\tablecaption{Summary of the three stacking methods studied in this paper.
\label{tab:methods}}
\tablehead{
\colhead{Stacking method} & \colhead{Sub-voxel centering?} & \colhead{Spatial profile} & \colhead{Spectral profile}}
\startdata
{\bf Original} & None & Not assumed & Not assumed \\
{\bf 2D forced photometry} & Spatial only & Assumed Gaussian matched to instrument beam or extended width& Not assumed \\
\raggedright{\bf 3D PRF fitting}\linebreak (`Gaussian', `Lorentzian') & Spatial and spectral & Assumed Gaussian matched to instrument beam or extended width& Assumed Gaussian or Lorentzian matched to model prediction \\
\enddata
\tablecomments{When we discuss `Gaussian' or `Lorentzian' 3D PRF fits, the distinction applies only to the spectral profile, and we always assume a Gaussian spatial profile.}\vspace{-5mm}
\end{deluxetable*}

\autoref{fig:StackSynopsis Placeholder} outlines the stacking methodologies described above based on 2D and 3D PRF models, in comparison to the original stack. \autoref{tab:methods} provides a written summary for additional clarity.

We may also conceive of fitting the 2D and 3D PRF models to the final average stacked cubelet from the original stack, rather than fitting per cutout. We find that this discards the advantage from sub-voxel centering and provides no sensitivity benefit, but we present results from sanity checks in~\autoref{sec:poststackappendix}.

\section{Basic verification with a toy model of sources on a uniform grid}\label{sec:sanitycheck}

\begin{deluxetable}{lc}
\tablecaption{Parameters of the simulated CO LIM observations.
\label{tab:simparams}}
\tablehead{
\colhead{Parameter} & \colhead{Value}}
\startdata
Survey field size & $4^\circ\times4^\circ$ \\
Pixel size & $2'\times2'$\\
Frequency range & 26--34 GHz \\
Voxel frequency width & 31.25 MHz \\
\enddata
\tablecomments{The simulated field size encompasses effectively the entire survey area of the COMAP Pathfinder, which in reality is split across three distinct patches of sky.}
\end{deluxetable}


For comparison with more robust simulations and eventually real data, we test both 2D and 3D PRF fitting methods against a simple toy model simulation of sources placed on a regular lattice. This allows us to verify that in the limit of sources sufficiently separated for zero confusion, the PRF stacks successfully recover the input toy model luminosity and are optimal when the PRF matches the toy model profile.

We generate 2250 sources uniformly distributed across the comoving volume, all with a luminosity of $6.86\times10^6\,L_\odot$ or $1.4\times10^{11}\,$K\,km\,s$^{-1}$\,pc$^2$. We arbitrarily cut down the mock source catalog to 2000 sources to be used in the stack (consistent with later simulations in~\autoref{sec:realisticsims}). We place these sources on a uniform 3D lattice with angular and spectral spacings of $7.5'$ and 727 MHz, before offsetting sources in the spectral dimension in increments of 62.5 MHz by as much as 438 MHz in either direction to avoid overlap in the line profiles of nearby sources. We then apply subvoxel angular and spectral offsets to all source positions, drawn from a uniform distribution. We generate a simulated CO map from these sources at $5\times$ spatial and spectral oversampling, convolving the luminosity density map with a 3D Gaussian profile of $5'$ angular FWHM (for a COMAP-like angular profile) and $62.5$ MHz spectral FWHM, before reducing to the final pixel size and frequency resolution. We describe the final map parameters in~\autoref{tab:simparams}.

The PRFs chosen span three spatial widths: $5'$, $9'$, and $20'$. A $5'$ width was chosen to match the width of the simulated COMAP-like profile; As an extreme comparison, a $20'$ width represents $1/12$ of the simulated survey field size and approximately $1/6$ of each field observed by the real-world COMAP Pathfinder survey. The $9'$ width acts as an intermediate value. For the 3D PRF, Gaussian and Lorentzian shapes were used along the spectral axis, in both cases with a fixed spectral FWHM of 2 channels (62.5 MHz).

We inject random Gaussian noise and explore a range of noise rms values per voxel spanning $\sigma\in(10^{-2},10^3)\,\mu$K. For comparison, if the luminosity of one of our toy model sources were completely undiluted and contained within a data cube voxel, we would expect each occupied voxel to have a brightness temperature of $\sim10^2\,\mu$K. Since the angular and spectral dilution spreads this luminosity across $\sim10^1$ voxels, we expect the map to achieve a SNR per occupied voxel of $\sim1$ at $\sigma\sim10^1\,\mu$K.

\begin{figure}
    \centering
    \begin{tikzpicture}
        \node () at (0,0) {\includegraphics[width=0.96\linewidth]{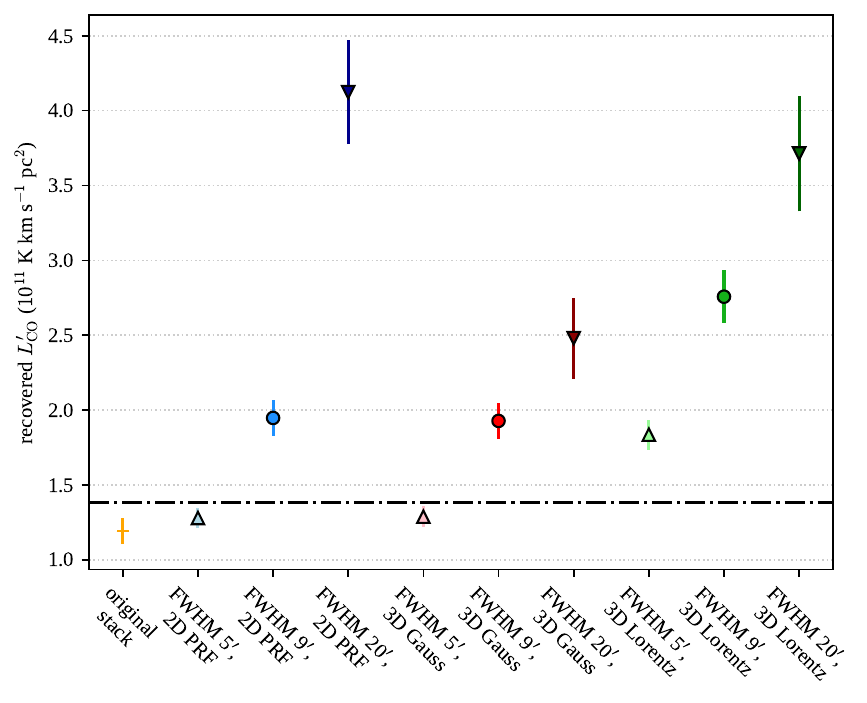}};
        \node () at (-2,2.5) {\Large\bf\emph{toy model}};
    \node () at (-2,1.86) {\large$\sigma=10\,\mu$K};
        \end{tikzpicture}
    \caption{Average \lprimeco{} recovered by PRF and original stacks from the toy model simulations of~\autoref{sec:sanitycheck}, with a simulated noise rms of $\sigma=10\,\mu$K.}
    \label{fig:PlaceholderGridRel5}
\end{figure}

We are able to confirm that in these simulations, the PRF methods recover the correct luminosity with no significant systematic biases if the correct profile is used. We find slight under-recovery ($L'\approx(1.28\pm0.07)\times10^{11}\,L_\odot$ for $\sigma=10\,\mu$K) compared to the input luminosity but attribute this to numerical errors (from imperfect positioning of sources in the $5\times$ oversampled but still discrete grid), as opposed to any inherent bias in the technique. In fact, the 3D Gaussian and 2D PRF stacks result in less biased recovery compared to the original stack ($L'\approx(1.19\pm0.09)\times10^{11}\,L_\odot$ for $\sigma=10\,\mu$K) , which misses flux outside the stack aperture. We show this graphically in~\autoref{fig:PlaceholderGridRel5}. Mismatched beam widths do result in systematic biases, with larger assumed beams resulting in larger apparently recovered flux.  (In the most extreme case, the ${\rm FWHM}=20'$ 2D PRF stack returns $L'\approx(4.1\pm0.3)\times10^{11}\,L_\odot$ for $\sigma=10\,\mu$K.) Similarly, as our ground truth spectral profile is Gaussian, using a Lorentzian spectral profile in our 3D PRF results in overestimation of flux, although the SNR is unaffected ($L'\approx(1.84\pm0.10)\times10^{11}\,L_\odot$ for $\sigma=10\,\mu$K). In fact, in these sanity checks we find the SNR to be predominantly affected by the FWHM of the angular profile, not by the choice of 2D or 3D PRF. Nonetheless, we find that not only does a well-matched profile allow for unbiased recovery, but it is also necessary for maximum SNR.

\begin{figure*}
    \centering
    \includegraphics[width=\linewidth]{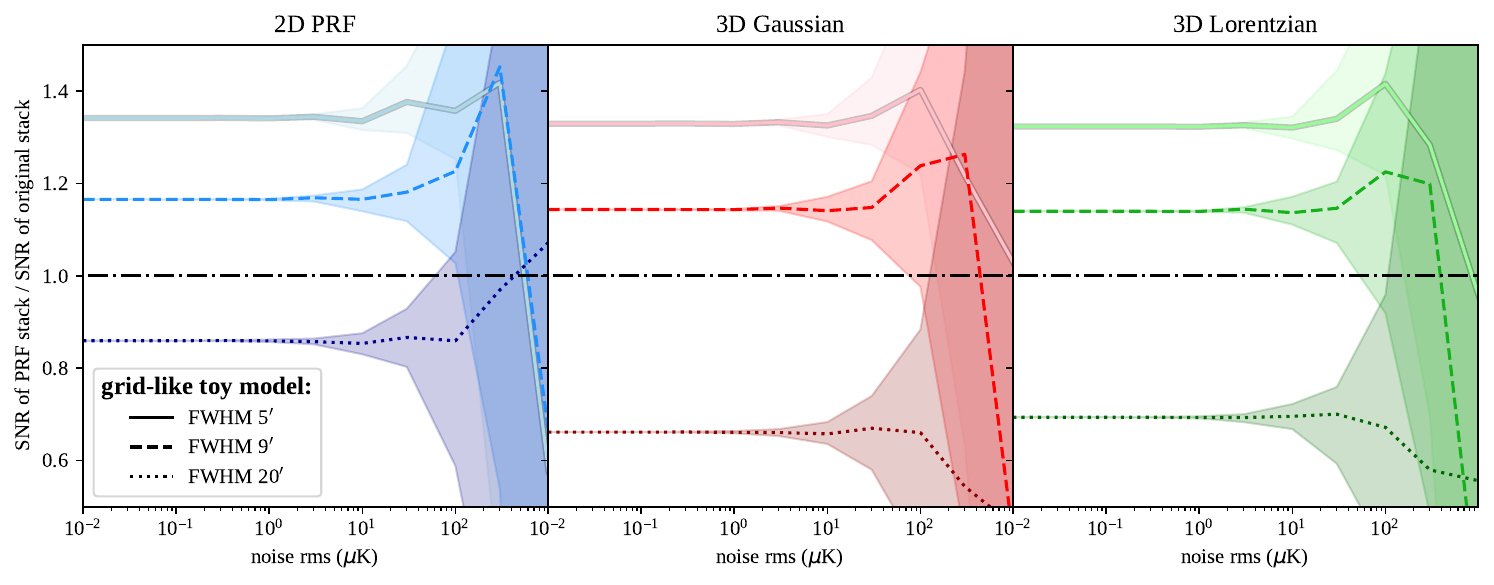}
    \caption{Ratio of SNR recovered by PRF stacking methods with different widths to the SNR of the original ($3\times5\times5$) stack across noise levels, using the toy model simulations of~\autoref{sec:sanitycheck} using a grid-like arrangement of sources with non-overlapping Gaussian emission profiles. The curves and shaded areas show median and 68\% interval values across 50 simulations of ratios of SNR for the stack method indicated by the line style and panel title. A horizontal line (black, dash-dotted) marks a ratio of 1.}
    \label{fig:PlaceholderGrid5}
\end{figure*}

\autoref{fig:PlaceholderGrid5} illustrates the SNR found in our toy model simulations. We find consistent behavior among all stacking methods up until noise-dominated regimes. We find that the PRF stacks recover a higher SNR than the original stack, with the optimal width being that of the input toy model profile (matching the COMAP beam width of $5'$), consistent with the signal spreading of point sources by the instrument beam. When fixing this angular FWHM, a PRF stack recovers 33\% more SNR than the original stack, again regardless of the exact 2D or 3D PRF approach. The increased SNR relative to the original stack comes principally from a $\sim20\%$ reduction in uncertainty rather than increased flux recovery, meaning we would expect upper limits to also improve by $\sim20$\% as long as we found the assumptions underlying the PRF model to be credible. PRF stacks with $9'$ angular FWHM still outperform the original stack by 14--17\%, but when the FWHM deviates too far away from ground truth to $20'$, the PRF stacks begin performing worse than the original stack by 14--33\%.

We ran a similar simulation with the convolved source profile in the simulated maps having an angular FWHM of $9'$ instead. We correspondingly found that the optimal PRF stack would use an angular FWHM of $9'$ as well, still outperforming the original stack in SNR by 14--17\% while the other PRF choices perform similarly to the original stack or under-perform by $\sim10$\% for a PRF choice of $20'$ angular FWHM. In this case, however, the increased SNR comes from cancellation of a $\sim70$\% improvement in flux recovery against a $\sim40\%$ increase in uncertainty. Therefore, upper limits would be worse at face value compared to the original stack, but not relative to the expected flux from any given model, and a detection would be more confident.

PRF stacking with both 2D and 3D PRFs thus shows the expected advantages over the original stacking with specific caveats. Our toy model simulations operate in the limit of perfectly separated, perfectly known emission profiles per source. But in this limit, the PRF model best matched to ground truth both performs best out of all our stacking pipelines and recovers the ground truth luminosity successfully. With this understanding in hand, we are ready to investigate what happens to PRF stacking outside this limit, when sources cluster and their emission profiles become confused.

\section{Realistic Tests: Results from Simulations with Clustered Sources} \label{sec:realisticsims}

To better understand the stack behavior on actual LIM data, we generate realistic simulations in which sources cluster together according to cosmological large-scale structure. We can then compare our findings in these simulations to behavior on the predictable maps of~\autoref{sec:sanitycheck}, to better understand what influences the efficacy of our stacking methods.

We generate a suite of simulations designed to realistically mock LIM observations with the \verb|joint_limlam_mocker|\footnote{\label{note1}\url{https://github.com/delaneydunne/joint_limlam_mocker}} code, following \cite{dunne2025_stacktheory}. These are based on peak-patch approximate cosmological simulations \citep{bondmeyers1996_peakpatchsims, stein2019_peakpatchsims} painted simultaneously with luminosities for CO and \lya{} emission lines based on the simulated dark matter halo masses. In this way, realistic interactions between the two tracers on the level of the large-scale structure are recovered. Mock LIM data cubes are then generated using the simulated CO emission, and mock galaxy catalogs with the simulated \lya\ emission. Several observational parameters (for example, the beam size, astrophysical line broadening, and the detection limit for the catalog) are also included in the simulation. We primarily follow the simulation parameters established in \cite{dunne2025_stacktheory}, with CO spectral line widths generated based on the expected circular velocity given halo mass (described, e.g., in \citealt{behroozi2019_universemachine}), yielding an average line width of $385\ \mathrm{km\ s^{-1}}$ per source.

One change from the simulations of~\cite{dunne2025_stacktheory} is that we continue to assume a beam FWHM of $5'$. Although the nominal beam FWHM for the COMAP Pathfinder is $4.5'$, the actual frequency- and feed-dependent beam FWHM ranges from $4.4'$ to $4.9'$, so a $5'$ Gaussian profile is sufficiently `COMAP-like' for the purposes of this work. These stacking methods and simulations are widely adaptable across a range of FWHMs and one close in width to that of the instrument is a sufficient proof of concept. We otherwise refer the reader to \cite{dunne2025_stacktheory} for a full description of the simulation process and the parameters used.

Using these simulations, we carry out the same analysis as in~\autoref{sec:sanitycheck} to investigate flux recovery and improvement in SNR by taking the flux and uncertainty estimates of our stacking pipeline at face value. The 3D PRFs still force a spectral FWHM of 62.5 MHz, as it is well matched to the stack profile FWHM of 596 km s$^{-1}$ found by~\cite{dunne2025_stacktheory} for simulations created in the same way as in this work.

\begin{figure}
    \centering
    \begin{tikzpicture}
    \node () at (0,0) {\includegraphics[width=0.96\linewidth]{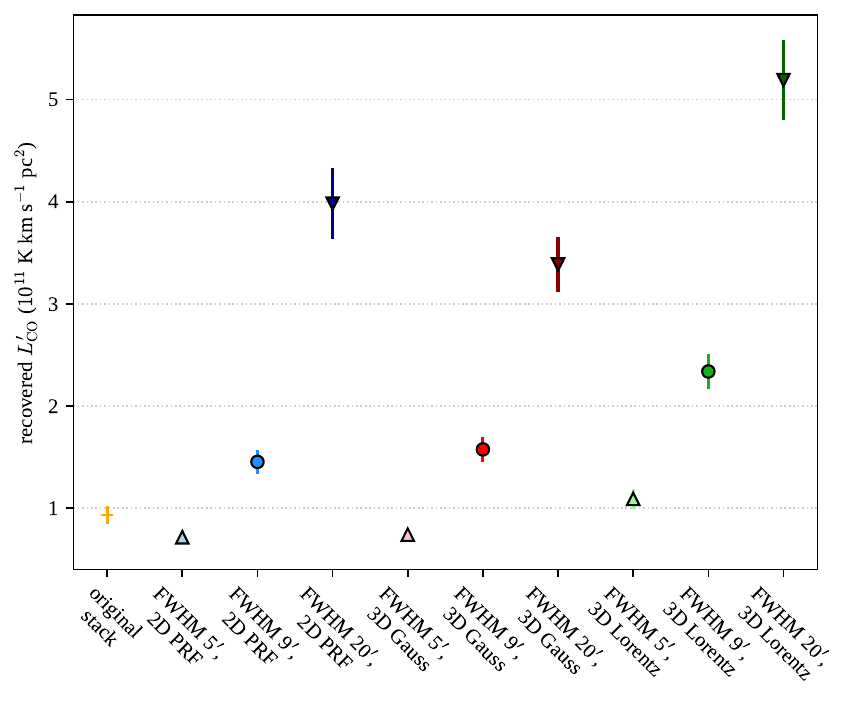}};
    \node () at (-1.086,2.586) {\Large\bf\emph{realistic simulations}};
    \node () at (-2,1.9586) {\large$\sigma=10\,\mu$K};
    \end{tikzpicture}
    \caption{Average \lprimeco{} recovered by PRF and original stacks from the realistic simulations of~\autoref{sec:realisticsims}, with a simulated noise rms of $\sigma=10\,\mu$K.}
    \label{fig:PlaceholderRealRel}
\end{figure}

\begin{figure*}
    \centering
    \includegraphics[width=\linewidth]{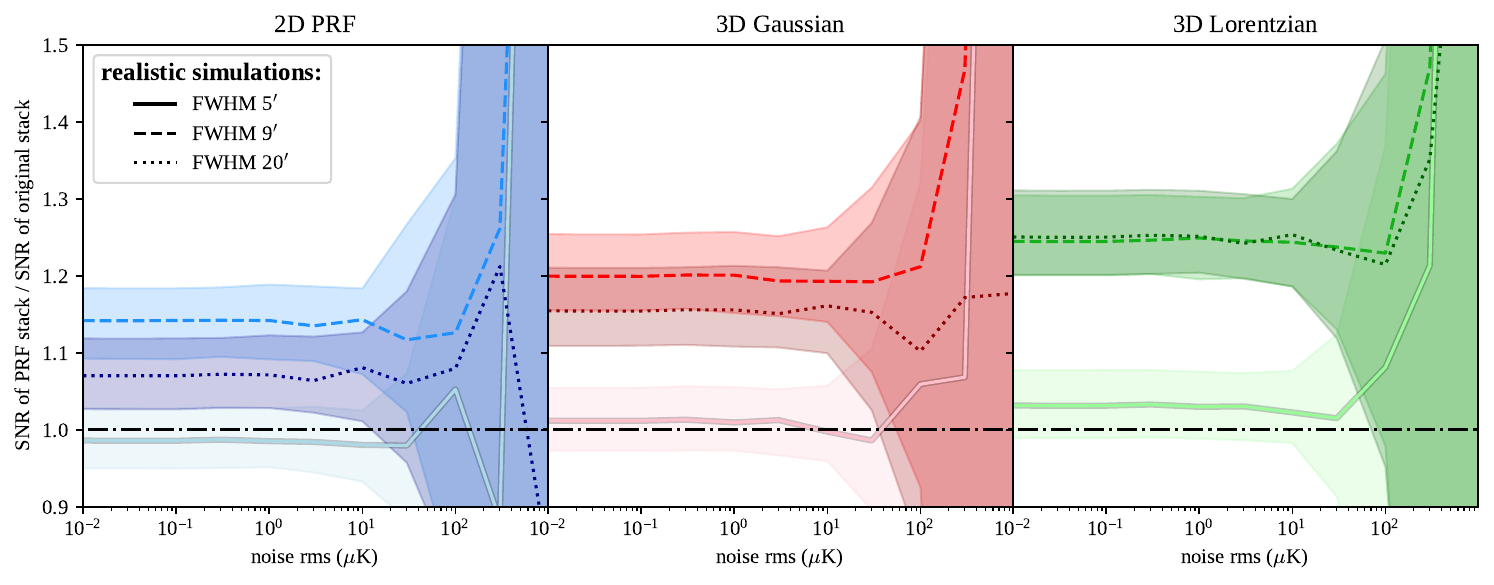}
    \caption{ Ratio of SNR recovered by PRF stacking methods with different widths to the SNR of the original (3 × 5 × 5) stack across noise
levels (similar to ~\autoref{fig:PlaceholderGrid5}), using the realistic simulations of~\autoref{sec:realisticsims}. The curves and shaded areas show median and 68\% interval values across 100 simulations of ratios of SNR for the stack method indicated by the
line style and panel title. A horizontal line (black, dash-dotted) marks a ratio of 1.}

    \label{fig:PlaceholderRealsim}
\end{figure*}

We assume instrument noise levels ranging from $10^{-2}$ $\mu$K to $10^3$ $\mu$K per voxel.\footnote{For reference, in COMAP Season 2 the Pathfinder survey achieves 25--50 $\mu$K noise per $2'\times2'\times31.25$ MHz voxel after three years of observations with 19 detectors on a single dish~\citep{lunde2024_COMAPS2_PaperI}.}  Across most of this range the instrument noise contributes most of the stack statistical variance, rather than sample variance. So we generate 100 semi-independent simulations, with the same underlying signal and sources but randomly re-generated noise and catalog selection.

Each of these 100 semi-independent simulations correspond to a COMAP-like survey field, still using the final map parameters of~\autoref{tab:simparams}, with 2000 sources randomly identified for each simulation out of all sources above a Ly$\alpha$ luminosity of $3\times10^{43}\,\mathrm{erg\,s^{-1}}$. We used the same 2D and 3D PRF models as in~\autoref{sec:sanitycheck}, spanning the same angular profile sizes ($5'$, $9'$, and $20'$), and in the case of the 3D PRF both Gaussian and Lorentzian spectral shapes (again keeping the spectral FWHM fixed at two frequency channels).

Figures \ref{fig:PlaceholderRealRel} and \ref{fig:PlaceholderRealsim} show, for all PRF models explored, the \lprimeco{} recovery for a noise level of $\sigma=10\,\mu$K and the SNR ratio achieved across different noise levels on these simulations. The profiles with a $5'$ FWHM result in roughly the same SNR as the original stack, whereas the best SNRs are recovered at a FWHM of $9'$ for the 2D and 3D Gaussian PRFs, and either $9'$ or $20'$ for the 3D Lorentzian PRF. Table \ref{tab:PH_RealSNRs} provides the average SNR advantages for each fitted profile for noise levels of 10 $\mu$K and 30 $\mu$K, the simulated levels most closely resembling the sensitivity of near-future real-world COMAP data.

\begin{table*}[]
    \centering
    \begin{tabular}{rcccccc}\hline\hline
    &\multicolumn{3}{c}{$\sigma=10\,\mu$K}&\multicolumn{3}{c}{$\sigma=30\,\mu$K}\\\cmidrule(lr){2-4}\cmidrule(lr){5-7}
          PRF angular FWHM & 2D&3D Gaussian&  3D Lorentzian& 2D&3D Gaussian&  3D Lorentzian\\\hline
          5$'$& $-2^{+4}_{-5}$\% & $+0^{+4}_{-6}$\% & $+2^{+4}_{-5}$\% & $-2^{+15}_{-12}$\% & $-1^{+14}_{-13}$\% & $+2^{+13}_{-12}$\%
    \\
 9$'$& $+14^{+6}_{-4}$\% & $+19^{+5}_{-5}$\% & $+24^{+5}_{-6}$\% & $+12^{+17}_{-15}$\% & $+19^{+15}_{-15}$\% & $+24^{+16}_{-15}$\%\\
 20$'$&$+8^{+3}_{-3}$\% & $+16^{+4}_{-5}$\% & $+25^{+5}_{-6}$\% & $+6^{+6}_{-6}$\% & $+15^{+13}_{-13}$\% & $+23^{+15}_{-12}$\%\\\hline
 \end{tabular}
    \caption{Median and 68\% sample interval across 100 realistic COMAP-like simulations of SNR advantages (disadvantages if negative) for PRF stacks relative to the original stack at the near-term attainable noise levels of $\sigma=10\,\mu$K and $\sigma=30\,\mu$K. Median percentages do not change dramatically for lower noise levels, although the interval narrows by a factor of 2--3.}
    \label{tab:PH_RealSNRs}
\end{table*}

For very high noise however, we clearly cannot reliably expect any PRF-based advantage in SNR over the original stack. Even when using the optimal PRF (FWHM $20'$, 3D, spectral Lorentzian), 25\% of simulations show worse SNR than the original stack for $\sigma=100\,\mu$K. For $\sigma<30\,\mu$K the percentage is less than 1\%, whereas for  $\sigma=300\,\mu$K this  rises to 58\% of simulations. This is purely a consequence of the fitted profile's inability to recover meaningful signal from each cutout. The absolute $\sigma$ value here is less meaningful than the per-voxel SNR; when our simulations reach per-voxel SNR values of $\lesssim0.1$, the PRF stack provides little reliable advantage. At that point, any signal is completely buried beneath noise and thus unable to be recovered with any method.

\section{Discussion}
\label{sec:discussion}
We must discuss two points to fully grasp the implications of the analysis carried out in this work. First, in~\autoref{sec:bootstraps}, we undertake a bootstrapping analysis on a downscaled stacking scenario in order to verify our understanding of uncertainties. Second, in~\autoref{sec:resultsdiscussion}, we further discuss how the SNR advantages found from PRF stacks in~\autoref{sec:realisticsims} highlights the peculiar nature of LIM stacking in comparison to conventional photometric surveys.
\subsection{Stack Uncertainties: Verification, Complications, and Interpretations}
\label{sec:bootstraps}
We aim to understand the uncertainties output from the pipeline for each stacking method using a bootstrapping analysis, following \cite{dunne2024comapearlyscienceviii}. For each target methodology, we perform 1000 randomized stacks. We use the toy model simulated maps generated in Section \ref{sec:sanitycheck}, and stack on a different catalog of randomized source positions for each bootstrap realization. We draw the random catalog positions from a uniform distribution spanning the entire map in all three dimensions. In this way, we can test the propagated uncertainties statistically. 

\begin{figure*}
    \centering
    \includegraphics[width=1\linewidth]{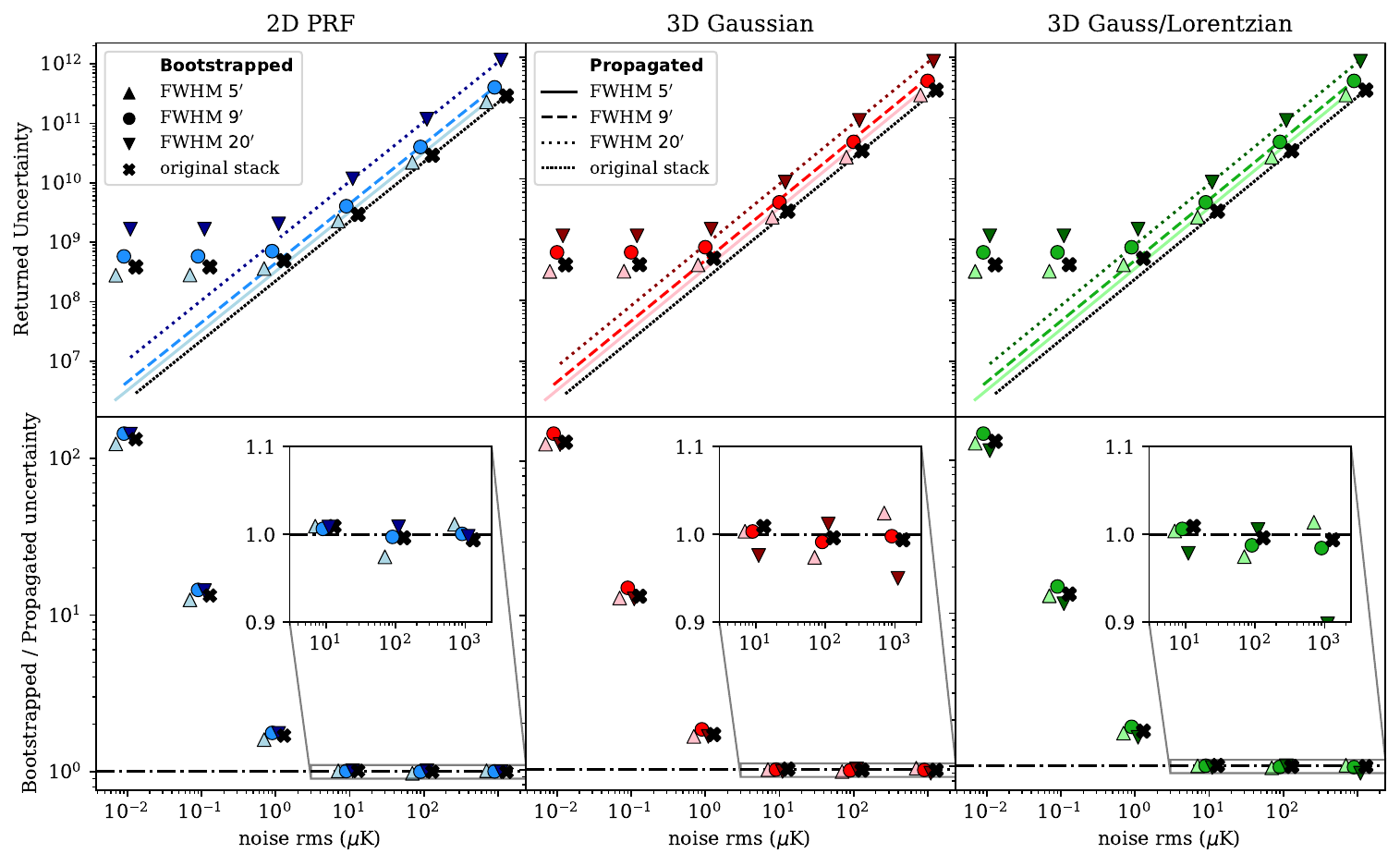}
    \caption{Uncertainties returned from error propagation compared to the uncertainties from bootstrapping for each stacking method. Points are slightly offset in the x-axes for clarity \textit{Top:} Returned uncertainties as a function of map RMS noise, for each of the three stacking methods. Points indicate bootstrapped uncertainties, and lines propagated uncertainties. The original stack (using a $5\times5\times3$ aperture) is shown in grey for comparison. \textit{Bottom:} Ratios between bootstrapped and propagated uncertainties. Where the map RMS dominates the uncertainty (above $\sim 10\,\mu$K), the propagated uncertainties agree well with the bootstraps. The inset axes show that the ratio between bootstrapped and propagated uncertainties is nearly 1 for all permutations. Below $\sim 10\,\mu$K, variance in the signal itself (reflected only in the bootstraps) begins to dominate.\label{fig:bootstraps}}
\end{figure*}

We compare the bootstrapped uncertainties to those output from the stacking pipelines for each of the three methodologies described in Section \ref{sec:stackingmethods}. We follow all other stack parameter choices made in Section \ref{sec:sanitycheck}: we test a range of map noise levels between $10^{-2}$ and $10^3\,\mu$K (which spans both high-significance simulations and those with more realistic noise levels), and we test three different profile widths in the spatial axes for each method (profile FWHM = $5'$, $9'$, and $20'$, to verify that the statistical uncertainties in the bootstraps match those output by the pipelines even across different PRF widths). The results of this comparison are shown in~\autoref{fig:bootstraps}. 

We find that, in the higher noise regimes, the uncertainty recovered from the bootstrapped error analysis is well-matched to the propagated uncertainties across all methods and PRF sizes, suggesting that the propagated errors (which are used throughout the rest of this work) are accurately reflecting all sources of statistical uncertainty. The uncertainty scales linearly with the map RMS and increases with the size of the PRF, as more voxels and thus more noise are being averaged into the stack. For all three stacking methodologies, the uncertainties of the `original' stack are slightly greater than the FWHM = $5'$ stacks, and less than the FWHM = $9'$ or $20'$ stacks. This is a function of the effective number of voxels of the stack cubelet summed into the luminosity measured for the stack -- the square $5\times5$ spatial aperture of the original stacking methodology (corresponding to $10'\times 10'$) incorporates more spatial pixels, and thus more map noise, than the Gaussian profile with a $5'$ FWHM. However, the Gaussian profile with a $9'$ FWHM has a core which is roughly the same size as the $5\times 5$ original stack, but also extends farther into the tails, and the $20'$ Gaussian profile is even larger.

While the two methods are consistent in the high-noise regime, the bootstrapped and propagated uncertainties begin to diverge at lower levels of map RMS noise ($\lesssim 1\,\mu$K). The propagated uncertainties continue to scale linearly with the map RMS, but the bootstrapped uncertainties flatten out, reaching a plateau of $3.8\times10^8\,\mathrm{K\,km\,s^{-1}\,pc^2}$ for the $5\times5\times3$ classic stack. This is true for all methodologies, and the ratio between the uncertainties returned for different methods and different beam sizes remains constant. This is because the injected random noise no longer dominates over sample variance -- the bootstraps are being performed on simulated maps with real (although simplistic) simulated signal, and the randomized catalogs are averaging in positions falling variously on and off of the simulated sources. In the high-noise regime, the signal level is below the level of the map RMS, so the sample variance isn't visible, but below $\sim 1\,\mu$K the sample variance begins to dominate. As the propagated uncertainties only account for the map-level RMS, and not the variance across the map, they continue to decrease linearly with decreasing map RMS.

\subsection{Response Functions Extended by Redshift-space Large-scale Structure}
\label{sec:resultsdiscussion}

Both of the stacking methodologies developed in this work are sensitive to the three-dimensional distribution of the stacked signal. Recent work in both 21 cm LIM~\citep{Chen25MeerKLASS} and CO LIM~\citep{dunne2025_stacktheory} show that the profile of a LIM stack on external source catalogs cannot be assumed to be simply the response function of the line-intensity field to a single point source. Understanding this fact is key to understanding our results. Therefore we take a brief moment to explain this expectation that at least at the spectral and spatial resolution typical of COMAP, the majority of the signal in the stack will be from large-scale cosmological clustering around the objects in the galaxy catalog, rather than the objects themselves. 

As~\cite{lunde2024_COMAPS2_PaperI} notes, voxels of $2'\times2'\times31.25$ MHz as used in this work correspond to a comoving volume of $3.7\times3.7\times4.1$ Mpc$^3$; the COMAP angular beam itself has a FWHM of around 9 Mpc in comoving space. By design, these scales correspond to resolving not individual sources but the signature of clustering. The fiducial CO model used in this work and in~\cite{dunne2025_stacktheory} comes from a suite of models derived by~\cite{chung2021_comapforecasts} from empirical constraints, all of which predict that the large-scale clustering of line emissivity becomes subdominant to the shot noise of discrete CO emiters at $k\sim0.2$--$0.3\,$Mpc$^{-1}$. This $k$ range corresponds to comoving features of $\pi/k\sim10$--$15$ Mpc, or around 2--4 voxel widths. As such, the profile of the stack emission, which we survey over multiple voxel widths in each dimension, cannot be well described by the line-intensity field's response to the presence of an individual line emitter. That means that the stack signal is a) not well-described by a single Gaussian distribution; b) more extended than the main beam alone in the spatial axes, or the astrophysical line broadening of a single galaxy in the spectral axis; and c) model-dependent, changing with the assumed distribution of both the CO and \lya\ luminosity overlaid upon the dark matter distribution. 

\begin{figure}
    \centering
    \includegraphics[width=0.986\linewidth]{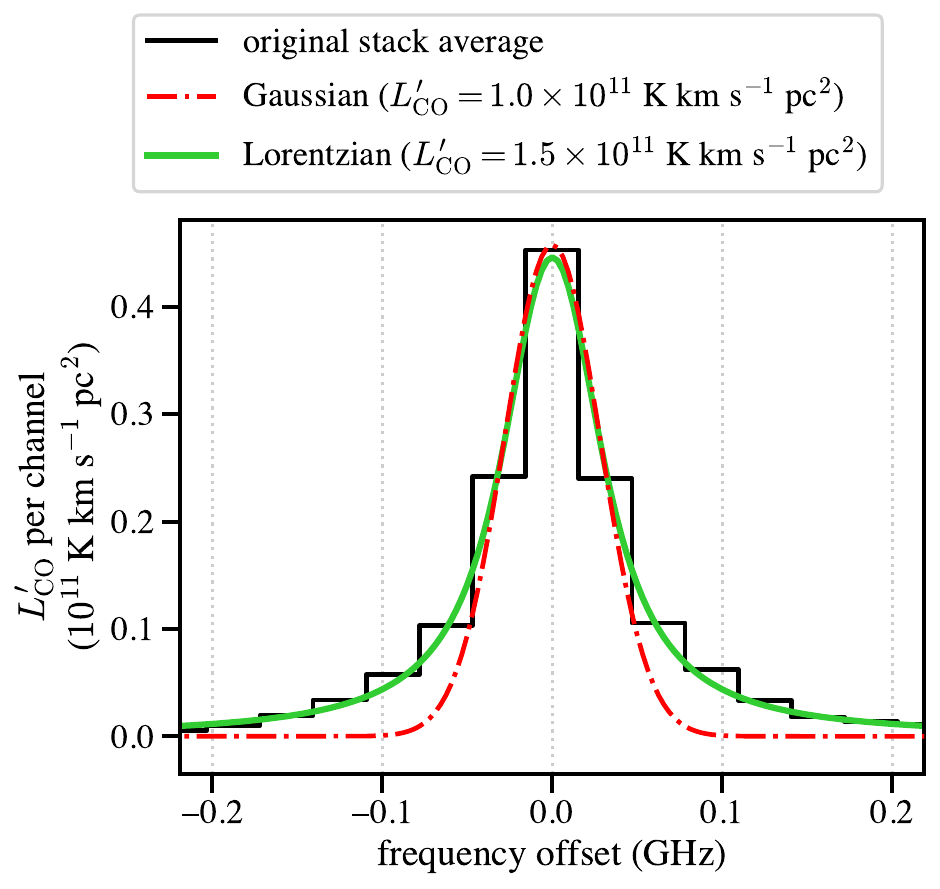}
    \caption{Comparison of the average spectrum of the original stack against best-fitting Gaussian and Lorentzian functions, with recovered total \lprimeco{} values shown in the legend.}
    \label{fig:profileillust}
\end{figure}

A stark illustration of this fact comes in the form of the average spectral profile of our simulated stack and the optimal choice of 3D PRF. The `Lorentzian' 3D PRF, with its heavy tails in the spectral dimension, always estimates larger integrated flux for a given line-intensity peak near the source position compared to a purely Gaussian 3D PRF. In our toy model of~\autoref{sec:sanitycheck}, where the ground truth is a purely Gaussian 3D response per source, this leads to a systematic bias in recovery of \lprimeco{} and a degraded SNR when using the Lorentzian profile. However, in the realistic simulations, the SNR is optimal for this Lorentzian profile. This indicates not only that the partly Lorentzian PRF provides a closer match to ground truth than a purely Gaussian 3D PRF, but also that the additional flux is real and it is the Gaussian PRF that is systematically biased. This is despite the fact that per source, the ground truth is a purely Gaussian profile.

For further insight on this point, we extract a spectrum from the average of the 100 original stacks calculated and compare the best-fitting Gaussian and Lorentzian spectral profiles in~\autoref{fig:profileillust}. The Gaussian PRF only captures the central peak, which is almost entirely contained in the central three channels.\footnote{As a side note, since both the original and 2D PRF stacking pipelines restrict themselves to summing luminosities over the central three channels,~\autoref{fig:profileillust} also explains their diminished flux recovery compared to even the 3D Gaussian PRF stacking pipeline. Extending the spectral aperture of the 2D PRF stack would likely improve flux recovery, but without any guide rails to assume correlations between the peak and tail, the SNR would diminish in a way similar to the original stack as examined in~\cite{dunne2025_stacktheory}.} However, with the same forced spectral FWHM of $\approx600$ km s$^{-1}$, the Lorentzian profile recovers 50\% more flux than the Gaussian by capturing the tails of the spectral profile.

What we see at work in~\autoref{fig:profileillust} is at least in part the fingers-of-God (FoG) damping of clustering in redshift space. As briefly noted in~\autoref{sec:realisticsims}, the average CO line width for each source does not exceed 400 km s$^{-1}$. The additional dispersion in the stack profile must come from the redshift-space distribution of CO emitters around source positions---including the random velocities with which the CO emitters move within large-scale overdensities, i.e., the FoG effect. Previous studies~\citep[e.g.:][]{2019ApJ...872..126M,anisotropies,2019PhRvD.100l3522B} have considered the FoG effect in varying ways, considering phenomenological damping functions in Fourier space, as well as dispersions ranging from 70 km s$^{-1}$ or 0.9 Mpc at $z\sim2.8$ in~\cite{anisotropies} (based on the study of~\citealt{Taruya2010}), to 250 km s$^{-1}$ in~\cite{2019ApJ...872..126M}, to 7 Mpc in~\cite{2019PhRvD.100l3522B}. What we see in~\autoref{fig:profileillust} is at the general scales expected of the FoG effect. However, a Lorentzian smearing of the two-point correlation function would correspond to a Laplace (double exponential) damping function applied in Fourier space, as opposed to the Gaussian and Lorentzian Fourier-space forms typically used most successfully in phenomenological descriptions of the FoG effect. The Lorentzian configuration-space profile is therefore likely a manifestation of different redshift-space clustering effects convolved together, as opposed to purely the FoG effect. This complexity poses a challenge for analytic models and thus strongly motivates continued investigation in simulations to improve comparisons against future LIM data.

As for the spatial dimensions, \cite{dunne2025_stacktheory} found that the angular profile of a COMAP-like stack is reasonably well-fit by the sum of two Gaussian profiles. For the fiducial COMAP modelling parameters (which are also the parameters assumed by the simulation work here), the standard deviations of the two best-fit Gaussian components are $3.2' \pm 0.1'$ and $9.6'\pm0.5'$, without deconvolving out the beam. These respectively correspond to FWHM values of $\approx8'$ and $\approx23'$, within 15\% or so of our extended PRF FWHM values of $9'$ and $20'$.


All of this perfectly explains the stark difference in stack behavior between our toy model simulations and realistic simulations for both 2D and 3D forced photometry. With a map of completely confusion-free, distinct line emission profiles, the stacking pipeline would maximize SNR when the fit profile matches the profile of the COMAP beam, as we saw in~\autoref{sec:sanitycheck} and specifically~\autoref{fig:PlaceholderGrid5}. However, when applied to realistically clustered, clumped-up sources that have overlapping signal contributions, we find that this changes with wider PRF models resulting in higher recovered SNRs. Therefore, we expect a better fit from wider PRFs to realistic data, and find that the luminosity values recovered depend heavily on the parameters used when fitting models. 

The results here also suggest that an incorrect choice of model parameters can result in a SNR no better than (or even worse than) that of the original stack. Any disadvantage is much less pronounced in the realistic case versus the toy model simulations, as we realistically always expect line emissivity to extend beyond cataloged sources. Still, this emphasizes the importance of carefully choosing signal profiles when stacking. This choice is unfortunately not trivial, as the profile depends on the underlying astrophysics that determine the luminosity function of line sources and the tracer bias of line emission. We discussed above that our extended PRF angular FWHM values match the Gaussian components quite well, and thus are optimal for the stacking pipeline as augmented in this work. But this optimality is only guaranteed under this set of CO and LAE model parameters. In the study by~\cite{dunne2025_stacktheory}, the spatial width of these components can change by $5$--$10\%$ with different models of CO emission, and up to $30\%$ if different galaxy populations are assumed. We have already also discussed complications regarding the interpretation of the stack spectral width, which in turn translate into complications for attempting to prescribe an optimal profile spectral width.

\section{Conclusions}
\label{sec:conclusions}
Our work provides basic answers to the questions we posed in \hyperref[sec:intro]{the Introduction}:
\begin{itemize}
    \item \emph{Can the forced fitting of 2D and 3D profiles at the sub-pixel level enhance the accuracy and significance of existing LIM stacking methods?} Yes. For our fiducial CO model, we find 2D and 3D PRF fitting begins to achieve a reliable SNR advantage over the original stack at noise levels being reached by the COMAP Pathfinder at the present time.
    \item \emph{Do the optimal parameters for these profiles depend solely on the instrument and survey design, or do they also depend on the LIM signal itself?} In a COMAP-like LIM survey, they depend strongly on details of the LIM signal that we can only anticipate phenomenologically through forward models at the present time.
\end{itemize}

The positive returns in the recovered flux and SNR when compared to the classic stack are not significant increases, but they are nonetheless non-trivial and notable. We see, on average, a 14--25\% increase when using extended PRF models in realistically generated simulations, the latter being equivalent to approximately 5--9 months of additional integration time over a 3-year COMAP dataset.

The necessity of using profiles that go beyond the point source response of the instrument itself reinforces one of the key insights of~\cite{dunne2025_stacktheory}: a cutout of LIM data contains a more complex signal distribution than a point source response function can anticipate. This is reflected in the behaviors between our realistic and toy model simulations. As discussed in~\autoref{sec:resultsdiscussion}, the extent of the profile in redshift space may even hold cosmological information about the large-scale redshift-space clustering and dispersion of line emission.


While we have discussed the advantages and complications of the PRF approach to stacking for LIM observations at COMAP-like resolution, these stacking methods and the takeaways we outline here may also prove useful for other LIM experiments in their early stages. That said, the ability to make these detections is ultimately limited by the raw sensitivity per voxel of the LIM experiment. Given more integration time, COMAP and other operating LIM experiments will achieve higher SNR and allow the refinements to methodology described in this work to further enhance sensitivity to signals.


In addition to the comparisons described in this work there are a number of remaining questions about these stacking methods which may hold valuable information. Our results suggest value in further study of stack outputs when using different PRF scales. In particular, stacks based on PRF models of different scales should essentially be different windows into the same line--galaxy two-point correlation function, and understanding this should bring a more coherent analytic picture behind these PRF stacks.

The biggest unexplored variable in the present work is the spectral profile shape and FWHM of the 3D PRF models. We only tested Gaussian and Lorentzian models with a FWHM of 2 channels, and CO line widths and clustering at high redshift are not so well understood that we can fix such parameters comfortably. The full spectro-photometric profile will also diverge significantly from the idealized profiles considered here owing to pipeline transfer functions. We leave consideration of all this to future work that will explore the PRF stack as a matched filter detection pipeline for the two-point correlation function at different scales, as well as the best ways to explore parameter variations in the stacking pipeline with less forced (spectro-)photometry.

Other ambitious yet natural extensions of this work would take catalog information into account beyond the position of one source at a time. A variation on oriented stacking (e.g.,~\citealt{Lokken25}), which uses the local orientation of the Cosmic Web to identify anisotropic matter distributions, could employ anisotropic PRF models aligned with the expected orientation of CO filaments. We are also exploring potential LIM applications of simultaneous stacking using fitting of PRF-convolved catalog hit maps to intensity data, as used in the context of cosmic infrared background emission from dusty star-forming galaxies~\citep[e.g.:][]{Viero13,Viero22}, another instance where clustering brings source confusion and affects photometric techniques.

\begin{acknowledgements}
DAD acknowledges support through the National Science Foundation grants AST-2206834 and AST-2406627. The authors thank members of the COMAP collaboration, including Kieran Cleary and Patrick Breysse, for their input into this study and for providing welcoming avenues for discussion. We also thank Kieran Cleary, Ben Vaughan, and Selina Yang for providing comments on an earlier draft of this manuscript.
\end{acknowledgements}

\software{\texttt{numpy}~\citep{numpy}; \texttt{scipy}~\citep{2020SciPy-NMeth}; \texttt{Matplotlib}~\citep{matplotlib}. This research made use of \texttt{Photutils}, an Astropy package for
detection and photometry of astronomical sources ~\citep{larry_bradley_2025_14889440}.}

\begin{contribution}

EMM designed and implemented the PRF stacking pipelines, oversaw the design of PRF and simulation parameters, and supervised the visualization of results from simulations. DAD implemented the bootstrap uncertainty verification, contributed the realistic LIM and catalog mocks, and provided much of the framework for the 2D PRF stack. DTC supervised the overall work and oversaw the execution of more detailed simulations, as well as contributed more advanced interpretation relating extended stack profiles to redshift-space clustering. All authors contributed equally to the design of the study, to the interpretation of results, and to the preparation and writing of the manuscript.

\end{contribution}

\appendix
\section{Post-stack Model Fitting}
\label{sec:poststackappendix}
\begin{figure}
    \centering
    \includegraphics[width=0.986\linewidth]{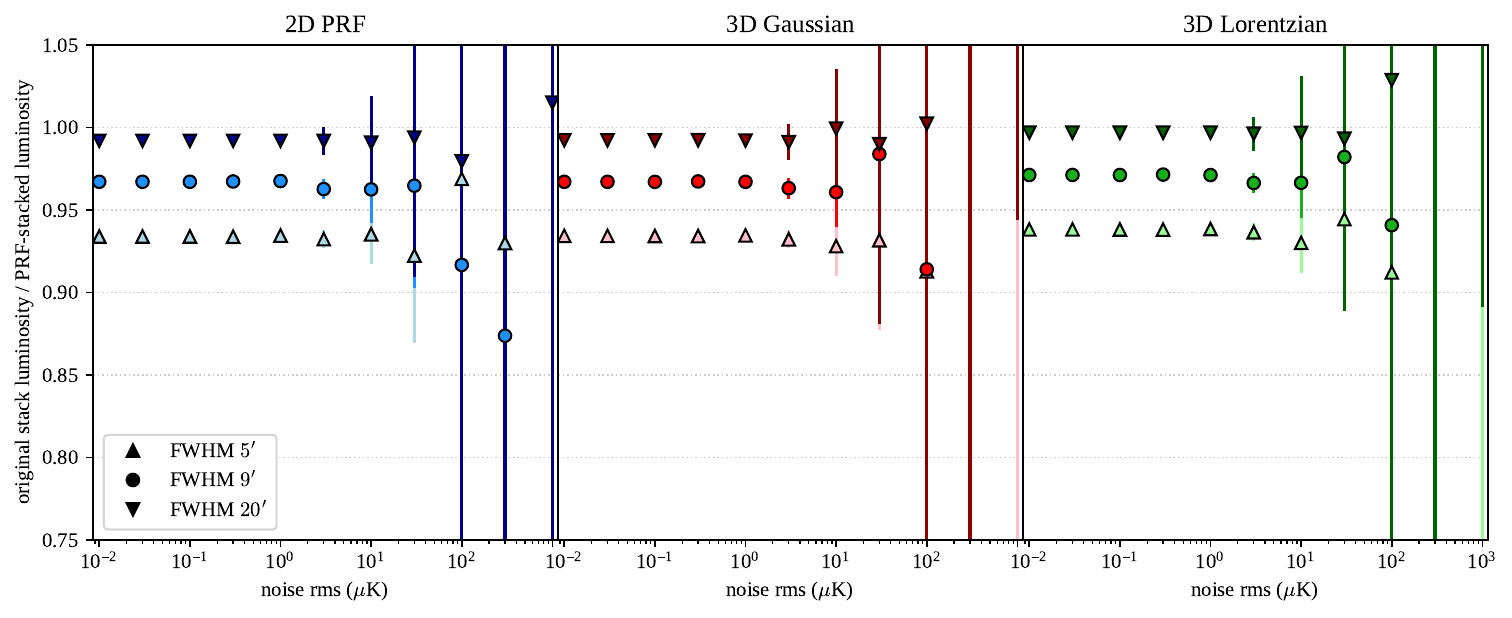}
    \caption{Ratio of \lprimeco{} values recovered by post-stack PRF fitting to those recovered by the PRF stacks described in the main text. We show the median and 68\% sample interval across 50 simulations using the markers and error bars across different simulated noise levels.}
    \label{fig:poststack}
\end{figure}

In this section, we briefly consider a post-processing of the original stack where the 2D and 3D PRF models are fit once to the final stacked cubelet, as opposed to the cutout-by-cutout PRF stacking method described in~\autoref{sec:stackingmethods} that we use in the remainder of this work. In this case, the center of the profile used in model fitting is taken as the center coordinate of the central voxel/pixel of the cubelet produced by the original stack. We test each method using the same simulation described in~\autoref{sec:sanitycheck}, where the sources are distributed in a grid-like fashion and convolved with a $5'$ FWHM angular Gaussian profile (mimicking the COMAP Pathfinder beam) and 62.5\,MHz FWHM spectral Gaussian profile. We show the results in~\autoref{fig:poststack}.

We find no advantage and in fact a slight disadvantage to using PRF fitting as a post-processing step for the original stack, compared to the iterative version we present above. The lack of adaptive sub-pixel/sub-voxel centering causes up to 6\% signal loss. As the stacking operation is entirely linear, we expect and find no advantage in uncertainty for PRF-based post-processing of the original stack over PRF stacking. Therefore, the pixel window smearing of the stack signal in the original stack causes signal loss and therefore can only disadvantage the SNR of PRF fitting to the original stack compared to PRF stacking.

\bibliography{bib}

@ARTICLE{Kovetz17,
       author = {{Kovetz}, Ely D. and {Viero}, Marco P. and {Lidz}, Adam and {Newburgh}, Laura and {Rahman}, Mubdi and {Switzer}, Eric and {Kamionkowski}, Marc and {Aguirre}, James and {Alvarez}, Marcelo and {Bock}, James and {Bond}, J. Richard and {Bower}, Goeffry and {Bradford}, C. Matt and {Breysse}, Patrick C. and {Bull}, Philip and {Chang}, Tzu-Ching and {Cheng}, Yun-Ting and {Chung}, Dongwoo and {Cleary}, Kieran and {Corray}, Asantha and {Crites}, Abigail and {Croft}, Rupert and {Dor{\'e}}, Olivier and {Eastwood}, Michael and {Ferrara}, Andrea and {Fonseca}, Jos{\'e} and {Jacobs}, Daniel and {Keating}, Garrett K. and {Lagache}, Guilaine and {Lakhlani}, Gunjan and {Liu}, Adrian and {Moodley}, Kavilan and {Murray}, Norm and {P{\'e}nin}, Aur{\'e}lie and {Popping}, Gerg{\"o} and {Pullen}, Anthony and {Reichers}, Dominik and {Saito}, Shun and {Saliwanchik}, Ben and {Santos}, Mario and {Somerville}, Rachel and {Stacey}, Gordon and {Stein}, George and {Villaescusa-Navarro}, Francesco and {Visbal}, Eli and {Weltman}, Amanda and {Wolz}, Laura and {Zemcov}, Micheal},
        title = "{Line-Intensity Mapping: 2017 Status Report}",
      journal = {arXiv e-prints},
         year = 2017,
        month = sep,
          eid = {arXiv:1709.09066},
        pages = {arXiv:1709.09066},
          doi = {10.48550/arXiv.1709.09066},
archivePrefix = {arXiv},
       eprint = {1709.09066},
 primaryClass = {astro-ph.CO},
       adsurl = {https://ui.adsabs.harvard.edu/abs/2017arXiv170909066K}
}

@ARTICLE{BernalKovetz22,
       author = {{Bernal}, Jos{\'e} Luis and {Kovetz}, Ely D.},
        title = "{Line-intensity mapping: theory review with a focus on star-formation lines}",
      journal = {\aapr},
         year = 2022,
        month = dec,
       volume = {30},
       number = {1},
          eid = {5},
        pages = {5},
          doi = {10.1007/s00159-022-00143-0},
archivePrefix = {arXiv},
       eprint = {2206.15377},
 primaryClass = {astro-ph.CO},
       adsurl = {https://ui.adsabs.harvard.edu/abs/2022A&ARv..30....5B}
}

@ARTICLE{dunne2025_stacktheory,
       author = {{Dunne}, D.~A. and {Cleary}, K.~A. and {Breysse}, P.~C. and {Chung}, D.~T. and {Ihle}, H.~T. and {Lunde}, J.~G.~S. and {Padmanabhan}, H. and {Stutzer}, N. -O. and {Bond}, J.~R. and {Gundersen}, J.~O. and {Kim}, J. and {Readhead}, A.~C.~S.},
        title = "{Three-Dimensional Stacking as a Line Intensity Mapping Statistic}",
      journal = {arXiv e-prints},
         year = 2025,
        month = mar,
          eid = {arXiv:2503.21743},
        pages = {arXiv:2503.21743},
          doi = {10.48550/arXiv.2503.21743},
archivePrefix = {arXiv},
       eprint = {2503.21743},
 primaryClass = {astro-ph.CO},
       adsurl = {https://ui.adsabs.harvard.edu/abs/2025arXiv250321743D}
}

@ARTICLE{stein2019_peakpatchsims,
       author = {{Stein}, George and {Alvarez}, Marcelo A. and {Bond}, J. Richard},
        title = "{The mass-Peak Patch algorithm for fast generation of deep all-sky dark matter halo catalogues and its N-body validation}",
      journal = {\mnras},
         year = 2019,
        month = feb,
       volume = {483},
       number = {2},
        pages = {2236-2250},
          doi = {10.1093/mnras/sty3226},
archivePrefix = {arXiv},
       eprint = {1810.07727},
 primaryClass = {astro-ph.CO},
       adsurl = {https://ui.adsabs.harvard.edu/abs/2019MNRAS.483.2236S}
}

@ARTICLE{bondmeyers1996_peakpatchsims,
       author = {{Bond}, J.~R. and {Myers}, S.~T.},
        title = "{The Peak-Patch Picture of Cosmic Catalogs. I. Algorithms}",
      journal = {\apjs},
         year = 1996,
        month = mar,
       volume = {103},
        pages = {1},
          doi = {10.1086/192267},
       adsurl = {https://ui.adsabs.harvard.edu/abs/1996ApJS..103....1B}
}

@ARTICLE{behroozi2019_universemachine,
       author = {{Behroozi}, Peter and {Wechsler}, Risa H. and {Hearin}, Andrew P. and {Conroy}, Charlie},
        title = "{UNIVERSEMACHINE: The correlation between galaxy growth and dark matter halo assembly from z = 0-10}",
      journal = {\mnras},
         year = 2019,
        month = sep,
       volume = {488},
       number = {3},
        pages = {3143-3194},
          doi = {10.1093/mnras/stz1182},
archivePrefix = {arXiv},
       eprint = {1806.07893},
 primaryClass = {astro-ph.GA},
       adsurl = {https://ui.adsabs.harvard.edu/abs/2019MNRAS.488.3143B}
}

@article{Cleary_2022,
   title={COMAP Early Science. I. Overview},
   volume={933},
   ISSN={1538-4357},
   url={http://dx.doi.org/10.3847/1538-4357/ac63cc},
   DOI={10.3847/1538-4357/ac63cc},
   number={2},
   journal={The Astrophysical Journal},
   publisher={American Astronomical Society},
   author={Cleary, Kieran A. and Borowska, Jowita and Breysse, Patrick C. and Catha, Morgan and Chung, Dongwoo T. and Church, Sarah E. and Dickinson, Clive and Eriksen, Hans Kristian and Foss, Marie Kristine and Gundersen, Joshua Ott and Harper, Stuart E. and Harris, Andrew I. and Hobbs, Richard and Ihle, Håvard T. and Kim, Junhan and Kocz, Jonathon and Lamb, James W. and Lunde, Jonas G. S. and Padmanabhan, Hamsa and Pearson, Timothy J. and Philip, Liju and Powell, Travis W. and Rasmussen, Maren and Readhead, Anthony C. S. and Rennie, Thomas J. and Silva, Marta B. and Stutzer, Nils-Ole and Uzgil, Bade D. and Watts, Duncan J. and Wehus, Ingunn Kathrine and Woody, David P. and Basoalto, Lilian and Bond, J. Richard and Dunne, Delaney A. and Gaier, Todd and Hensley, Brandon and Keating, Laura C. and Lawrence, Charles R. and Murray, Norman and Paladini, Roberta and Reeves, Rodrigo and Viero, Marco P. and Wechsler, Risa H.},
   year={2022},
   month=jul, pages={182} }

@article{Lamb_2022,
   title={COMAP Early Science. II. Pathfinder Instrument},
   volume={933},
   ISSN={1538-4357},
   url={http://dx.doi.org/10.3847/1538-4357/ac63c6},
   DOI={10.3847/1538-4357/ac63c6},
   number={2},
   journal={The Astrophysical Journal},
   publisher={American Astronomical Society},
   author={Lamb, James W. and Cleary, Kieran A. and Woody, David P. and Catha, Morgan and Chung, Dongwoo T. and Gundersen, Joshua Ott and Harper, Stuart E. and Harris, Andrew I. and Hobbs, Richard and Ihle, Håvard T. and Kocz, Jonathon and Pearson, Timothy J. and Philip, Liju and Powell, Travis W. and Basoalto, Lilian and Bond, J. Richard and Borowska, Jowita and Breysse, Patrick C. and Church, Sarah E. and Dickinson, Clive and Dunne, Delaney A. and Eriksen, Hans Kristian and Foss, Marie Kristine and Gaier, Todd and Kim, Junhan and Lawrence, Charles R. and Lunde, Jonas G. S. and Padmanabhan, Hamsa and Rasmussen, Maren and Readhead, Anthony C. S. and Reeves, Rodrigo and Rennie, Thomas J. and Stutzer, Nils-Ole and Viero, Marco P. and Watts, Duncan J. and Wehus, Ingunn Kathrine},
   year={2022},
   month=jul, pages={183} }

@misc{dunne2024comapearlyscienceviii,
      title={COMAP Early Science: VIII. A Joint Stacking Analysis with eBOSS Quasars}, 
      author={Delaney A. Dunne and Kieran A. Cleary and Patrick C. Breysse and Dongwoo T. Chung and Havard T. Ihle and J. Richard Bond and Hans Kristian Eriksen and Joshua Ott Gundersen and Laura C. Keating and Junhan Kim and Jonas Gahr Sturtzel Lunde and Norman Murray and Hamsa Padmanabhan and Liju Philip and Nils-Ole Stutzer and Doga Tolgay and Ingunn Katherine Wehus and Sarah E. Church and Todd Gaier and Andrew I. Harris and Richard Hobbs and James W. Lamb and Charles R. Lawrence and Anthony C. S. Readhead and David P. Woody},
      year={2024},
      eprint={2304.09832},
      archivePrefix={arXiv},
      primaryClass={astro-ph.GA},
      url={https://arxiv.org/abs/2304.09832}, 
}

@article{Dawson_2016,
   title={THE SDSS-IV EXTENDED BARYON OSCILLATION SPECTROSCOPIC SURVEY: OVERVIEW AND EARLY DATA},
   volume={151},
   ISSN={1538-3881},
   url={http://dx.doi.org/10.3847/0004-6256/151/2/44},
   DOI={10.3847/0004-6256/151/2/44},
   number={2},
   journal={The Astronomical Journal},
   publisher={American Astronomical Society},
   author={Dawson, Kyle S. and Kneib, Jean-Paul and Percival, Will J. and Alam, Shadab and Albareti, Franco D. and Anderson, Scott F. and Armengaud, Eric and Aubourg, Éric and Bailey, Stephen and Bautista, Julian E. and Berlind, Andreas A. and Bershady, Matthew A. and Beutler, Florian and Bizyaev, Dmitry and Blanton, Michael R. and Blomqvist, Michael and Bolton, Adam S. and Bovy, Jo and Brandt, W. N. and Brinkmann, Jon and Brownstein, Joel R. and Burtin, Etienne and Busca, N. G. and Cai, Zheng and Chuang, Chia-Hsun and Clerc, Nicolas and Comparat, Johan and Cope, Frances and Croft, Rupert A. C. and Cruz-Gonzalez, Irene and Costa, Luiz N. da and Cousinou, Marie-Claude and Darling, Jeremy and Macorra, Axel de la and Torre, Sylvain de la and Delubac, Timothée and Bourboux, Hélion du Mas des and Dwelly, Tom and Ealet, Anne and Eisenstein, Daniel J. and Eracleous, Michael and Escoffier, S. and Fan, Xiaohui and Finoguenov, Alexis and Font-Ribera, Andreu and Frinchaboy, Peter and Gaulme, Patrick and Georgakakis, Antonis and Green, Paul and Guo, Hong and Guy, Julien and Ho, Shirley and Holder, Diana and Huehnerhoff, Joe and Hutchinson, Timothy and Jing, Yipeng and Jullo, Eric and Kamble, Vikrant and Kinemuchi, Karen and Kirkby, David and Kitaura, Francisco-Shu and Klaene, Mark A. and Laher, Russ R. and Lang, Dustin and Laurent, Pierre and Goff, Jean-Marc Le and Li, Cheng and Liang, Yu and Lima, Marcos and Lin, Qiufan and Lin, Weipeng and Lin, Yen-Ting and Long, Daniel C. and Lundgren, Britt and MacDonald, Nicholas and Maia, Marcio Antonio Geimba and Malanushenko, Elena and Malanushenko, Viktor and Mariappan, Vivek and McBride, Cameron K. and McGreer, Ian D. and Ménard, Brice and Merloni, Andrea and Meza, Andres and Montero-Dorta, Antonio D. and Muna, Demitri and Myers, Adam D. and Nandra, Kirpal and Naugle, Tracy and Newman, Jeffrey A. and Noterdaeme, Pasquier and Nugent, Peter and Ogando, Ricardo and Olmstead, Matthew D. and Oravetz, Audrey and Oravetz, Daniel J. and Padmanabhan, Nikhil and Palanque-Delabrouille, Nathalie and Pan, Kaike and Parejko, John K. and Pâris, Isabelle and Peacock, John A. and Petitjean, Patrick and Pieri, Matthew M. and Pisani, Alice and Prada, Francisco and Prakash, Abhishek and Raichoor, Anand and Reid, Beth and Rich, James and Ridl, Jethro and Rodriguez-Torres, Sergio and Rosell, Aurelio Carnero and Ross, Ashley J. and Rossi, Graziano and Ruan, John and Salvato, Mara and Sayres, Conor and Schneider, Donald P. and Schlegel, David J. and Seljak, Uros and Seo, Hee-Jong and Sesar, Branimir and Shandera, Sarah and Shu, Yiping and Slosar, Anže and Sobreira, Flavia and Streblyanska, Alina and Suzuki, Nao and Taylor, Donna and Tao, Charling and Tinker, Jeremy L. and Tojeiro, Rita and Vargas-Magaña, Mariana and Wang, Yuting and Weaver, Benjamin A. and Weinberg, David H. and White, Martin and Wood-Vasey, W. M. and Yeche, Christophe and Zhai, Zhongxu and Zhao, Cheng and Zhao, Gong-bo and Zheng, Zheng and Zhu, Guangtun Ben and Zou, Hu},
   year={2016},
   month=feb, pages={44} }

@article{Mentuch_Cooper_2023,
   title={HETDEX Public Source Catalog 1: 220 K Sources Including Over 50 K Lyα Emitters from an Untargeted Wide-area Spectroscopic Survey*},
   volume={943},
   ISSN={1538-4357},
   url={http://dx.doi.org/10.3847/1538-4357/aca962},
   DOI={10.3847/1538-4357/aca962},
   number={2},
   journal={The Astrophysical Journal},
   publisher={American Astronomical Society},
   author={Mentuch Cooper, Erin and Gebhardt, Karl and Davis, Dustin and Farrow, Daniel J. and Liu, Chenxu and Zeimann, Gregory and Ciardullo, Robin and Feldmeier, John J. and Drory, Niv and Jeong, Donghui and Benda, Barbara and Bowman, William P. and Boylan-Kolchin, Michael and Chávez Ortiz, Óscar A. and Debski, Maya H. and Dentler, Mona and Fabricius, Maximilian and Farooq, Rameen and Finkelstein, Steven L. and Gawiser, Eric and Gronwall, Caryl and Hill, Gary J. and Hopp, Ulrich and House, Lindsay R. and Janowiecki, Steven and Khoraminezhad, Hasti and Kollatschny, Wolfram and Komatsu, Eiichiro and Landriau, Martin and Niemeyer, Maja Lujan and Lee, Hanshin and MacQueen, Phillip and Mawatari, Ken and McKay, Brianna and Ouchi, Masami and Poppe, Jennifer and Saito, Shun and Schneider, Donald P. and Snigula, Jan and Thomas, Benjamin P. and Tuttle, Sarah and Urrutia, Tanya and Weiss, Laurel and Wisotzki, Lutz and Zhang, Yechi},
   year={2023},
   month=feb, pages={177} }

@software{larry_bradley_2025_14889440,
  author       = {Larry Bradley and
                  Brigitta Sipőcz and
                  Thomas Robitaille and
                  Erik Tollerud and
                  Zé Vinícius and
                  Christoph Deil and
                  Kyle Barbary and
                  Tom J Wilson and
                  Ivo Busko and
                  Axel Donath and
                  Hans Moritz Günther and
                  Mihai Cara and
                  P. L. Lim and
                  Sebastian Meßlinger and
                  Zach Burnett and
                  Simon Conseil and
                  Michael Droettboom and
                  Azalee Bostroem and
                  E. M. Bray and
                  Lars Andersen Bratholm and
                  William Jamieson and
                  Adam Ginsburg and
                  Geert Barentsen and
                  Matt Craig and
                  Sergio Pascual and
                  Shivangee Rathi and
                  Marshall Perrin and
                  Brett M. Morris},
  title        = {astropy/photutils: 2.2.0},
  month        = feb,
  year         = 2025,
  publisher    = {Zenodo},
  version      = {2.2.0},
  doi          = {10.5281/zenodo.14889440},
  url          = {https://doi.org/10.5281/zenodo.14889440},
  swhid        = {swh:1:dir:11159107f27a28985192ed1118b1f2055709d093
                   ;origin=https://doi.org/10.5281/zenodo.596036;visi
                   t=swh:1:snp:ae8c4a55d349d43e53cfe9ce92e678fcfe840f
                   3b;anchor=swh:1:rel:0117f67e8888adcdfc85308287dd9c
                   854b466389;path=astropy-photutils-ffb96c5
                  },
}

@ARTICLE{2020SciPy-NMeth,
  author  = {Virtanen, Pauli and Gommers, Ralf and Oliphant, Travis E. and
            Haberland, Matt and Reddy, Tyler and Cournapeau, David and
            Burovski, Evgeni and Peterson, Pearu and Weckesser, Warren and
            Bright, Jonathan and {van der Walt}, St{\'e}fan J. and
            Brett, Matthew and Wilson, Joshua and Millman, K. Jarrod and
            Mayorov, Nikolay and Nelson, Andrew R. J. and Jones, Eric and
            Kern, Robert and Larson, Eric and Carey, C J and
            Polat, {\.I}lhan and Feng, Yu and Moore, Eric W. and
            {VanderPlas}, Jake and Laxalde, Denis and Perktold, Josef and
            Cimrman, Robert and Henriksen, Ian and Quintero, E. A. and
            Harris, Charles R. and Archibald, Anne M. and
            Ribeiro, Ant{\^o}nio H. and Pedregosa, Fabian and
            {van Mulbregt}, Paul and {SciPy 1.0 Contributors}},
  title   = {{{SciPy} 1.0: Fundamental Algorithms for Scientific
            Computing in Python}},
  journal = {Nature Methods},
  year    = {2020},
  volume  = {17},
  pages   = {261--272},
  adsurl  = {https://rdcu.be/b08Wh},
  doi     = {10.1038/s41592-019-0686-2},
}

@ARTICLE{dawson2016_ebossoverview,
       author = {{Dawson}, Kyle S. and {Kneib}, Jean-Paul and {Percival}, Will J. and {Alam}, Shadab and {Albareti}, Franco D. and {Anderson}, Scott F. and {Armengaud}, Eric and {Aubourg}, {\'E}ric and {Bailey}, Stephen and {Bautista}, Julian E. and {Berlind}, Andreas A. and {Bershady}, Matthew A. and {Beutler}, Florian and {Bizyaev}, Dmitry and {Blanton}, Michael R. and {Blomqvist}, Michael and {Bolton}, Adam S. and {Bovy}, Jo and {Brandt}, W.~N. and {Brinkmann}, Jon and {Brownstein}, Joel R. and {Burtin}, Etienne and {Busca}, N.~G. and {Cai}, Zheng and {Chuang}, Chia-Hsun and {Clerc}, Nicolas and {Comparat}, Johan and {Cope}, Frances and {Croft}, Rupert A.~C. and {Cruz-Gonzalez}, Irene et al},
        title = "{The SDSS-IV Extended Baryon Oscillation Spectroscopic Survey: Overview and Early Data}",
      journal = {\apj},
         year = 2016,
        month = feb,
       volume = {151},
       number = {2},
          eid = {44},
        pages = {44},
          doi = {10.3847/0004-6256/151/2/44},
archivePrefix = {arXiv},
       eprint = {1508.04473},
 primaryClass = {astro-ph.CO},
       adsurl = {https://ui.adsabs.harvard.edu/abs/2016AJ....151...44D}
}

@ARTICLE{Yang25,
       author = {{Yang}, Selina F. and {McAtee}, Sophie M. and {Vaughan}, Benjamin J. and {Crites}, Abigail T. and {Butler}, Victoria L. and {Chung}, Dongwoo T. and {Keenan}, Ryan P. and {Pham}, Dang and {Prakash}, Shwetha and {Bock}, James J. and {Bradford}, Charles M. and {Chang}, Tzu-Ching and {Cheng}, Yun-Ting and {Dunn}, Audrey and {Emerson}, Nicholas and {Frez}, Clifford and {Hunacek}, Jonathon and {Li}, Chao-Te and {Lowe}, Ian N. and {Lau}, King and {Marrone}, Daniel P. and {Mayer}, Evan C. and {Sun}, Guochao and {Trumper}, Isaac and {Turner}, Anthony D. and {Wei}, Ta-Shun and {Zemcov}, Michael},
        title = "{TIME Commissioning Observations: I. Mapping Dust and Molecular Gas in the Sgr A Molecular Cloud Complex at the Galactic Center}",
      journal = {arXiv e-prints},
         year = 2025,
        month = nov,
          eid = {arXiv:2511.09473},
        pages = {arXiv:2511.09473},
          doi = {10.48550/arXiv.2511.09473},
archivePrefix = {arXiv},
       eprint = {2511.09473},
 primaryClass = {astro-ph.IM},
       adsurl = {https://ui.adsabs.harvard.edu/abs/2025arXiv251109473Y}
}

@article{CW13,
	Archiveprefix = {arXiv},
	Author = {{Carilli}, C.~L. and {Walter}, F.},
	Doi = {10.1146/annurev-astro-082812-140953},
	Eprint = {1301.0371},
	Journal = {\araa},
	Month = aug,
	Pages = {105-161},
	Primaryclass = {astro-ph.CO},
	Title = {{Cool Gas in High-Redshift Galaxies}},
	Url = {http://adsabs.harvard.edu/abs/2013ARA%26A..51..105C},
	Volume = 51,
	Year = 2013}

@ARTICLE{Desert25,
       author = {{D{\'e}sert}, F.-X. and {Mac{\'\i}as-P{\'e}rez}, J.~F. and {Beelen}, A. and {Beno{\^\i}t}, A. and {B{\'e}thermin}, M. and {Bounmy}, J. and {Bourrion}, O. and {Calvo}, M. and {Catalano}, A. and {De Breuck}, C. and {Dubois}, C. and {Dur{\'a}n}, C.~A. and {Fasano}, A. and {Goupy}, J. and {Hu}, W. and {Ibar}, E. and {Lagache}, G. and {Lundgren}, A. and {Monfardini}, A. and {Ponthieu}, N. and {Quinatoa}, D. and {Van Cuyck}, M. and {Adam}, R. and {Ade}, P. and {Ajeddig}, H. and {Amarantidis}, S. and {Andr{\'e}}, P. and {Aussel}, H. and {Berta}, S. and {Bongiovanni}, A. and {Ch{\'e}rouvrier}, D. and {De Petris}, M. and {Doyle}, S. and {Driessen}, E.~F.~C. and {Ejlali}, G. and {Ferragamo}, A. and {Gomez}, A. and {Hanser}, C. and {Katsioli}, S. and {K{\'e}ruzor{\'e}}, F. and {Kramer}, C. and {Ladjelate}, B. and {Leclercq}, S. and {Lestrade}, J.-F. and {Madden}, S.~C. and {Maury}, A. and {Mayet}, F. and {Moyer-Anin}, A. and {Mu{\~n}oz-Echeverr{\'\i}a}, M. and {Myserlis}, I. and {Paliwal}, A. and {Perotto}, L. and {Pisano}, G. and {Rev{\'e}ret}, V. and {Rigby}, A.~J. and {Ritacco}, A. and {Roussel}, H. and {Ruppin}, F. and {S{\'a}nchez-Portal}, M. and {Savorgnano}, S. and {Sievers}, A. and {Tucker}, C. and {Zylka}, R.},
        title = "{Continuum, CO, and water vapour maps of the Orion Nebula: First millimetre spectral imaging with CONCERTO}",
      journal = {\aap},
         year = 2025,
        month = sep,
       volume = {701},
          eid = {A210},
        pages = {A210},
          doi = {10.1051/0004-6361/202555320},
archivePrefix = {arXiv},
       eprint = {2504.20487},
 primaryClass = {astro-ph.GA},
       adsurl = {https://ui.adsabs.harvard.edu/abs/2025A&A...701A.210D}
}

@article{Chang10,
	Author = {{Chang}, T.-C. and {Pen}, U.-L. and {Bandura}, K. and {Peterson}, J.~B.},
	Doip = {10.1038/nature09187},
	Journal = {\nat},
	Month = jul,
	Pages = {463-465},
	Title = {{An intensity map of hydrogen 21-cm emission at redshift z\~{}0.8}},
	Url = {http://adsabs.harvard.edu/abs/2010Natur.466..463C},
	Volume = 466,
	Year = 2010}

@ARTICLE{Masui13,
       author = {{Masui}, K.~W. and {Switzer}, E.~R. and {Banavar}, N. and {Bandura}, K.
        and {Blake}, C. and {Calin}, L. -M. and {Chang}, T. -C. and
        {Chen}, X. and {Li}, Y. -C. and {Liao}, Y. -W. and {Natarajan},
        A. and {Pen}, U. -L. and {Peterson}, J.~B. and {Shaw}, J.~R. and
        {Voytek}, T.~C.},
        title = "{Measurement of 21 cm Brightness Fluctuations at z \textasciitilde 0.8 in
        Cross-correlation}",
      journal = {\apj},
         year = 2013,
        month = Jan,
       volume = {763},
          eid = {L20},
        pages = {L20},
          doi = {10.1088/2041-8205/763/1/L20},
archivePrefix = {arXiv},
       eprint = {1208.0331},
 primaryClass = {astro-ph.CO},
       adsurl = {https://ui.adsabs.harvard.edu/#abs/2013ApJ...763L..20M}
}

@ARTICLE{CHIME2023,
       author = {{Amiri}, Mandana and {Bandura}, Kevin and {Chen}, Tianyue and {Deng}, Meiling and {Dobbs}, Matt and {Fandino}, Mateus and {Foreman}, Simon and {Halpern}, Mark and {Hill}, Alex S. and {Hinshaw}, Gary and {H{\"o}fer}, Carolin and {Kania}, Joseph and {Landecker}, T.~L. and {MacEachern}, Joshua and {Masui}, Kiyoshi and {Mena-Parra}, Juan and {Milutinovic}, Nikola and {Mirhosseini}, Arash and {Newburgh}, Laura and {Ordog}, Anna and {Pen}, Ue-Li and {Pinsonneault-Marotte}, Tristan and {Polzin}, Ava and {Reda}, Alex and {Renard}, Andre and {Shaw}, J. Richard and {Siegel}, Seth R. and {Singh}, Saurabh and {Vanderlinde}, Keith and {Wang}, Haochen and {Wiebe}, Donald V. and {Wulf}, Dallas and {CHIME Collaboration}},
        title = "{Detection of Cosmological 21 cm Emission with the Canadian Hydrogen Intensity Mapping Experiment}",
      journal = {\apj},
         year = 2023,
        month = apr,
       volume = {947},
       number = {1},
          eid = {16},
        pages = {16},
          doi = {10.3847/1538-4357/acb13f},
archivePrefix = {arXiv},
       eprint = {2202.01242},
 primaryClass = {astro-ph.CO},
       adsurl = {https://ui.adsabs.harvard.edu/abs/2023ApJ...947...16A}
}

@ARTICLE{comap_xcorr,
       author = {{Chung}, Dongwoo T. and {Viero}, Marco P. and {Church}, Sarah E. and
         {Wechsler}, Risa H. and {Alvarez}, Marcelo A. and {Bond}, J. Richard and
         {Breysse}, Patrick C. and {Cleary}, Kieran A. and {Eriksen}, Hans K. and
         {Foss}, Marie K. and {Gundersen}, Joshua O. and {Harper}, Stuart E. and
         {Ihle}, H{\r{a}}vard T. and {Keating}, Laura C. and {Murray}, Norman and
         {Padmanabhan}, Hamsa and {Stein}, George F. and {Wehus}, Ingunn K. and
         {COMAP Collaboration}},
        title = "{Cross-correlating Carbon Monoxide Line-intensity Maps with Spectroscopic and Photometric Galaxy Surveys}",
      journal = {\apj},
         year = "2019",
        month = "Feb",
       volume = {872},
       number = {2},
          eid = {186},
        pages = {186},
          doi = {10.3847/1538-4357/ab0027},
archivePrefix = {arXiv},
       eprint = {1809.04550},
 primaryClass = {astro-ph.GA},
       adsurl = {https://ui.adsabs.harvard.edu/abs/2019ApJ...872..186C}
}

@ARTICLE{Niemeyer25,
       author = {{Lujan Niemeyer}, Maja and {Komatsu}, Eiichiro and {Bernal}, Jos{\'e} Luis and {Byrohl}, Chris and {Ciardullo}, Robin and {Curtis}, Olivia and {Farrow}, Daniel J. and {Finkelstein}, Steven L. and {Gebhardt}, Karl and {Gronwall}, Caryl and {Hill}, Gary J. and {Jarvis}, Matt J. and {Jeong}, Donghui and {Mentuch Cooper}, Erin and {Mitra}, Deeshani and {Mukae}, Shiro and {Mu{\~n}oz}, Julian B. and {Ouchi}, Masami and {Saito}, Shun and {Schneider}, Donald P. and {Wisotzki}, Lutz},
        title = "{Ly{\ensuremath{\alpha}} Intensity Mapping in HETDEX: Galaxy-Ly{\ensuremath{\alpha}} Intensity Cross-Power Spectrum}",
      journal = {arXiv e-prints},
         year = 2025,
        month = oct,
          eid = {arXiv:2510.11427},
        pages = {arXiv:2510.11427},
          doi = {10.48550/arXiv.2510.11427},
archivePrefix = {arXiv},
       eprint = {2510.11427},
 primaryClass = {astro-ph.CO},
       adsurl = {https://ui.adsabs.harvard.edu/abs/2025arXiv251011427L}
}

@ARTICLE{Chen25MeerKLASS,
       author = {{Chen}, Zhaoting and {Cunnington}, Steven and {Pourtsidou}, Alkistis and {Wolz}, Laura and {Spinelli}, Marta and {Bernal}, Jos{\'e} Luis and {Barberi-Squarotti}, Matilde and {Camera}, Stefano and {Carucci}, Isabella P. and {Fonseca}, Jos{\'e} and {Grainge}, Keith and {Irfan}, Melis O. and {Santos}, Mario G. and {Wang}, Jingying and {Meerklass Collaboration}},
        title = "{Emission-line Stacking of 21 cm Intensity Maps with MeerKLASS: Inference Pipeline and Application to the L-band Deep-field Data}",
      journal = {\apjs},
         year = 2025,
        month = jul,
       volume = {279},
       number = {1},
          eid = {19},
        pages = {19},
          doi = {10.3847/1538-4365/add897},
archivePrefix = {arXiv},
       eprint = {2504.03908},
 primaryClass = {astro-ph.CO},
       adsurl = {https://ui.adsabs.harvard.edu/abs/2025ApJS..279...19C}
}

@ARTICLE{Taruya2010,
       author = {{Taruya}, Atsushi and {Nishimichi}, Takahiro and {Saito}, Shun},
        title = "{Baryon acoustic oscillations in 2D: Modeling redshift-space power spectrum from perturbation theory}",
      journal = {\prd},
         year = "2010",
        month = "Sep",
       volume = {82},
       number = {6},
          eid = {063522},
        pages = {063522},
          doi = {10.1103/PhysRevD.82.063522},
archivePrefix = {arXiv},
       eprint = {1006.0699},
 primaryClass = {astro-ph.CO},
       adsurl = {https://ui.adsabs.harvard.edu/abs/2010PhRvD..82f3522T}
}

@ARTICLE{anisotropies,
       author = {{Chung}, Dongwoo T.},
        title = "{A Partial Inventory of Observational Anisotropies in Single-dish Line-intensity Mapping}",
      journal = {\apj},
         year = "2019",
        month = "Aug",
       volume = {881},
       number = {2},
          eid = {149},
        pages = {149},
          doi = {10.3847/1538-4357/ab3040},
archivePrefix = {arXiv},
       eprint = {1905.00209},
 primaryClass = {astro-ph.CO},
       adsurl = {https://ui.adsabs.harvard.edu/abs/2019ApJ...881..149C}
}

@ARTICLE{2019PhRvD.100l3522B,
       author = {{Bernal}, Jos{\'e} Luis and {Breysse}, Patrick C. and {Gil-Mar{\'\i}n}, H{\'e}ctor and {Kovetz}, Ely D.},
        title = "{User's guide to extracting cosmological information from line-intensity maps}",
      journal = {\prd},
         year = 2019,
        month = dec,
       volume = {100},
       number = {12},
          eid = {123522},
        pages = {123522},
          doi = {10.1103/PhysRevD.100.123522},
archivePrefix = {arXiv},
       eprint = {1907.10067},
 primaryClass = {astro-ph.CO},
       adsurl = {https://ui.adsabs.harvard.edu/abs/2019PhRvD.100l3522B}
}

@ARTICLE{2019ApJ...872..126M,
       author = {{Moradinezhad Dizgah}, Azadeh and {Keating}, Garrett K.},
        title = "{Line Intensity Mapping with [C II] and CO(1-0) as Probes of Primordial Non-Gaussianity}",
      journal = {\apj},
         year = 2019,
        month = feb,
       volume = {872},
       number = {2},
          eid = {126},
        pages = {126},
          doi = {10.3847/1538-4357/aafd36},
archivePrefix = {arXiv},
       eprint = {1810.02850},
 primaryClass = {astro-ph.CO},
       adsurl = {https://ui.adsabs.harvard.edu/abs/2019ApJ...872..126M}
}

@ARTICLE{lunde2024_COMAPS2_PaperI,
       author = {{Lunde}, J.~G.~S. and {Stutzer}, N. -O. and {Breysse}, P.~C. and {Chung}, D.~T. and {Cleary}, K.~A. and {Dunne}, D.~A. and {Eriksen}, H.~K. and {Harper}, S.~E. and {Ihle}, H.~T. and {Lamb}, J.~W. and {Pearson}, T.~J. and {Philip}, L. and {Wehus}, I.~K. and {Woody}, D.~P. and {Bond}, J.~R. and {Church}, S.~E. and {Gaier}, T. and {Gundersen}, J.~O. and {Harris}, A.~I. and {Hobbs}, R. and {Kim}, J. and {Lawrence}, C.~R. and {Murray}, N. and {Padmanabhan}, H. and {Readhead}, A.~C.~S. and {Rennie}, T.~J. and {Tolgay}, D. and {COMAP Collaboration}},
        title = "{COMAP Pathfinder {\textendash} Season 2 results: I. Improved data selection and processing}",
      journal = {\aap},
         year = 2024,
        month = nov,
       volume = {691},
          eid = {A335},
        pages = {A335},
          doi = {10.1051/0004-6361/202451121},
archivePrefix = {arXiv},
       eprint = {2406.07510},
 primaryClass = {astro-ph.CO},
       adsurl = {https://ui.adsabs.harvard.edu/abs/2024A&A...691A.335L}
}

@ARTICLE{chung2021_comapforecasts,
       author = {{Chung}, Dongwoo T. and {Breysse}, Patrick C. and {Cleary}, Kieran A. and {Ihle}, H{\r{a}}vard T. and {Padmanabhan}, Hamsa and {Silva}, Marta B. and {Richard Bond}, J. and {Borowska}, Jowita and {Catha}, Morgan and {Church}, Sarah E. and {Dunne}, Delaney A. and {Kristian Eriksen}, Hans and {Kristine Foss}, Marie and {Gaier}, Todd and {Ott Gundersen}, Joshua and {Harper}, Stuart E. and {Harris}, Andrew I. and {Hensley}, Brandon and {Hobbs}, Richard and {Keating}, Laura C. and {Kim}, Junhan and {Lamb}, James W. and {Lawrence}, Charles R. and {Gahr Sturtzel Lunde}, Jonas and {Murray}, Norman and {Pearson}, Timothy J. and {Philip}, Liju and {Rasmussen}, Maren and {Readhead}, Anthony C.~S. and {Rennie}, Thomas J. and {Stutzer}, Nils-Ole and {Uzgil}, Bade D. and {Viero}, Marco P. and {Watts}, Duncan J. and {Wechsler}, Risa H. and {Kathrine Wehus}, Ingunn and {Woody}, David P. and {Comap Collaboration}},
        title = "{COMAP Early Science. V. Constraints and Forecasts at z   3}",
      journal = {\apj},
         year = 2022,
        month = jul,
       volume = {933},
       number = {2},
          eid = {186},
        pages = {186},
          doi = {10.3847/1538-4357/ac63c7},
archivePrefix = {arXiv},
       eprint = {2111.05931},
 primaryClass = {astro-ph.CO},
       adsurl = {https://ui.adsabs.harvard.edu/abs/2022ApJ...933..186C}
}

@ARTICLE{dunne2025_comaps2iv,
       author = {{Dunne}, D.~A. and {Cleary}, K.~A. and {Lunde}, J.~G.~S. and {Chung}, D.~T. and {Breysse}, P.~C. and {Stutzer}, N.~O. and {Bond}, J.~R. and {Eriksen}, H.~K. and {Gundersen}, J.~O. and {Hoerning}, G.~A. and {Kim}, J. and {Mansfield}, E.~M. and {Mason}, S.~R. and {Murray}, N. and {Rennie}, T.~J. and {Tolgay}, D. and {Valentine}, S. and {Wehus}, I.~K. and {COMAP Collaboration}},
        title = "{COMAP Pathfinder -- Season 2 results IV. A stack on eBOSS/DESI quasars}",
      journal = {arXiv e-prints},
         year = 2025,
        month = oct,
          eid = {arXiv:2510.23568},
        pages = {arXiv:2510.23568},
          doi = {10.48550/arXiv.2510.23568},
archivePrefix = {arXiv},
       eprint = {2510.23568},
 primaryClass = {astro-ph.CO},
       adsurl = {https://ui.adsabs.harvard.edu/abs/2025arXiv251023568D}
}

@Article{matplotlib,
  Author    = {Hunter, J. D.},
  Title     = {Matplotlib: A 2D graphics environment},
  Journal   = {Computing in Science \& Engineering},
  Volume    = {9},
  Number    = {3},
  Pages     = {90--95},
  publisher = {IEEE COMPUTER SOC},
  doi       = {10.1109/MCSE.2007.55},
  year      = 2007
}

@ARTICLE{Lokken25,
       author = {{Lokken}, M. and {van Engelen}, A. and {Aguena}, M. and {Allam}, S.~S. and {Anbajagane}, D. and {Bacon}, D. and {Baxter}, E. and {Blazek}, J. and {Bocquet}, S. and {Bond}, J.~R. and {Brooks}, D. and {Calabrese}, E. and {Carnero Rosell}, A. and {Carretero}, J. and {Costanzi}, M. and {da Costa}, L.~N. and {Coulton}, W.~R. and {De Vicente}, J. and {Desai}, S. and {Doel}, P. and {Doux}, C. and {Duivenvoorden}, A.~J. and {Dunkley}, J. and {Huang}, Z. and {Everett}, S. and {Ferrero}, I. and {Frieman}, J. and {Garc{\'\i}a-Bellido}, J. and {Gatti}, M. and {Gaztanaga}, E. and {Giannini}, G. and {Gluscevic}, Vera and {Gruen}, D. and {Gruendl}, R.~A. and {Guan}, Y. and {Gutierrez}, G. and {Hinton}, S.~R. and {Hlozek}, R. and {Hollowood}, D.~L. and {Honscheid}, K. and {James}, D.~J. and {Kuehn}, K. and {Lahav}, O. and {Lee}, S. and {Li}, Z. and {Madhavacheril}, M. and {Marques}, G.~A. and {Marshall}, J.~L. and {Mena-Fern{\'a}ndez}, J. and {Menanteau}, F. and {Miquel}, R. and {Myles}, J. and {Niemack}, M.~D. and {Pandey}, S. and {Pereira}, M.~E.~S. and {Pieres}, A. and {Plazas Malag{\'o}n}, A.~A. and {Porredon}, A. and {Rodr{\'\i}guez-Monroy}, M. and {Roodman}, A. and {Samuroff}, S. and {Sanchez}, E. and {Sanchez Cid}, D. and {Santiago}, B. and {Schubnell}, M. and {Sevilla-Noarbe}, I. and {Sif{\'o}n}, C. and {Smith}, M. and {Staggs}, S.~T. and {Suchyta}, E. and {Swanson}, M.~E.~C. and {Tarle}, G. and {To}, C.-H. and {Weaverdyck}, N. and {Wiseman}, P. and {Wollack}, E.~J.},
        title = "{Superclustering with the Atacama Cosmology Telescope and Dark Energy Survey. II. Anisotropic Large-scale Coherence in Hot Gas, Galaxies, and Dark Matter}",
      journal = {\apj},
         year = 2025,
        month = apr,
       volume = {982},
       number = {2},
          eid = {186},
        pages = {186},
          doi = {10.3847/1538-4357/adb622},
archivePrefix = {arXiv},
       eprint = {2409.04535},
 primaryClass = {astro-ph.CO},
       adsurl = {https://ui.adsabs.harvard.edu/abs/2025ApJ...982..186L}
}

@ARTICLE{Viero13,
       author = {{Viero}, M.~P. and {Moncelsi}, L. and {Quadri}, R.~F. and {Arumugam}, V. and {Assef}, R.~J. and {B{\'e}thermin}, M. and {Bock}, J. and {Bridge}, C. and {Casey}, C.~M. and {Conley}, A. and {Cooray}, A. and {Farrah}, D. and {Glenn}, J. and {Heinis}, S. and {Ibar}, E. and {Ikarashi}, S. and {Ivison}, R.~J. and {Kohno}, K. and {Marsden}, G. and {Oliver}, S.~J. and {Roseboom}, I.~G. and {Schulz}, B. and {Scott}, D. and {Serra}, P. and {Vaccari}, M. and {Vieira}, J.~D. and {Wang}, L. and {Wardlow}, J. and {Wilson}, G.~W. and {Yun}, M.~S. and {Zemcov}, M.},
        title = "{HerMES: The Contribution to the Cosmic Infrared Background from Galaxies Selected by Mass and Redshift}",
      journal = {\apj},
         year = 2013,
        month = dec,
       volume = {779},
       number = {1},
          eid = {32},
        pages = {32},
          doi = {10.1088/0004-637X/779/1/32},
archivePrefix = {arXiv},
       eprint = {1304.0446},
 primaryClass = {astro-ph.CO},
       adsurl = {https://ui.adsabs.harvard.edu/abs/2013ApJ...779...32V}
}

@ARTICLE{Viero22,
       author = {{Viero}, Marco P. and {Sun}, Guochao and {Chung}, Dongwoo T. and {Moncelsi}, Lorenzo and {Condon}, Sam S.},
        title = "{The early Universe was dust-rich and extremely hot}",
      journal = {\mnras},
         year = 2022,
        month = oct,
       volume = {516},
       number = {1},
        pages = {L30-L34},
          doi = {10.1093/mnrasl/slac075},
archivePrefix = {arXiv},
       eprint = {2203.14312},
 primaryClass = {astro-ph.GA},
       adsurl = {https://ui.adsabs.harvard.edu/abs/2022MNRAS.516L..30V}
}

@ARTICLE{numpy,
       author = {{Harris}, Charles R. and {Millman}, K. Jarrod and {van der Walt}, St{\'e}fan J. and {Gommers}, Ralf and {Virtanen}, Pauli and {Cournapeau}, David and {Wieser}, Eric and {Taylor}, Julian and {Berg}, Sebastian and {Smith}, Nathaniel J. and {Kern}, Robert and {Picus}, Matti and {Hoyer}, Stephan and {van Kerkwijk}, Marten H. and {Brett}, Matthew and {Haldane}, Allan and {del R{\'\i}o}, Jaime Fern{\'a}ndez and {Wiebe}, Mark and {Peterson}, Pearu and {G{\'e}rard-Marchant}, Pierre and {Sheppard}, Kevin and {Reddy}, Tyler and {Weckesser}, Warren and {Abbasi}, Hameer and {Gohlke}, Christoph and {Oliphant}, Travis E.},
        title = "{Array programming with NumPy}",
      journal = {\nat},
         year = 2020,
        month = sep,
       volume = {585},
       number = {7825},
        pages = {357-362},
          doi = {10.1038/s41586-020-2649-2},
archivePrefix = {arXiv},
       eprint = {2006.10256},
 primaryClass = {cs.MS},
       adsurl = {https://ui.adsabs.harvard.edu/abs/2020Natur.585..357H}
}
\bibliographystyle{aasjournalv7}

\end{document}